\documentclass[twocolumn]{aastex631}
\usepackage{sidecap}
\usepackage{amsmath}
\usepackage{ulem}
\usepackage{tabularx}
\DeclareUnicodeCharacter{0301}{\'{e}}

\begin{document}

\title{Measuring frequency and period separations in red-giant stars using machine learning}

\author{Siddharth Dhanpal}
\email{dhanpal.siddharth@gmail.com}
\affiliation{Department of Astronomy and Astrophysics, Tata Institute of Fundamental Research, Mumbai, 400005, India}

\author{Othman Benomar}
\affiliation{ Center for Space Science, NYUAD Institute, New York University Abu Dhabi, PO Box 129188, Abu Dhabi, UAE}
\affiliation{Division of Solar and Plasma Astrophysics, NAOJ, Mitaka, Tokyo, Japan}

\author{Shravan Hanasoge}
\affiliation{Department of Astronomy and Astrophysics, Tata Institute of Fundamental Research, Mumbai, 400005, India}
\affiliation{ Center for Space Science, NYUAD Institute, New York University Abu Dhabi, PO Box 129188, Abu Dhabi, UAE}

\author{Abhisek Kundu}
\affiliation{Parallel Computing Lab, Intel Labs, Bangalore, India}

\author{Dattaraj Dhuri}
\affiliation{Department of Astronomy and Astrophysics, Tata Institute of Fundamental Research, Mumbai, 400005, India}
 
\author{Dipankar Das}
\affiliation{Parallel Computing Lab, Intel Labs, Bangalore, India}

\author{Bharat Kaul}
\affiliation{Parallel Computing Lab, Intel Labs, Bangalore, India}

\begin{abstract}
Asteroseismology is used to infer the interior physics of stars. The \textit{Kepler} and TESS space missions have provided a vast data set of red-giant light curves, which may be used for asteroseismic analysis. These data sets are expected to significantly grow with future missions such as \textit{PLATO}, and efficient methods are therefore required to analyze these data rapidly. Here, we describe a machine learning algorithm that identifies red giants from the raw oscillation spectra and captures \textit{p} and \textit{mixed} mode parameters from the red-giant power spectra. We report algorithmic inferences for large frequency separation ($\Delta \nu$), frequency at maximum amplitude ($\nu_{max}$), and period separation ($\Delta \Pi$) for an ensemble of stars. In addition, we have discovered $\sim$25 new probable red giants among 151,000 \textit{Kepler} long-cadence stellar-oscillation spectra analyzed by the method, among which four are binary candidates which appear to possess red-giant counterparts. To validate the results of this method, we selected $\sim$ 3,000 \textit{Kepler} stars, at various evolutionary stages ranging from subgiants to red clumps, and compare inferences of $\Delta \nu$, $\Delta \Pi$, and $\nu_{max}$ with estimates obtained using other techniques. The power of the machine-learning algorithm lies in its speed: it is able to accurately extract seismic parameters from 1,000 spectra in $\sim$5 seconds on a modern computer\footnote{single core of the Intel\textsuperscript{\textregistered} Xeon\textsuperscript{\textregistered} Platinum 8280 CPU}.    
\end{abstract}

\section{Introduction}
\label{Introduction}
Asteroseismology is an important tool that sheds light on stellar physics, allowing us to understand their inner structure and  evolution. Space-borne instruments such as \textit{CoRoT} \citep{Baglin2006a}, \textit{Kepler} \citep{Borucki2004,Borucki2010}, and TESS \citep{Ricker2015} have observed hundreds of thousands of stars and detected stellar pulsations in tens of thousands of them. Most of these pulsating stars are red giants \citep{Mosser2010a,Yu_2018}. Red giants are evolved solar-like stars, and, as in the Sun, pulsations are driven by turbulence in the outer layers of the convection zone. In most of those stars, only a few seismic characteristics have been identified \citep{Bugnet2018,Hon2019}. 

Detailed studies of red giants have significantly improved our understanding of the interiors and evolution of stars \citep{Bedding2011,Mosser_2014}. It also helped in probing their rotation \citep{Beck2012,six_stars,Di_Mauro_2016}, showing potential indications for strong magnetic fields in their inner layers \citep{Fuller423}. Extending these detailed analyses to a broader range of stars (if not all solar-like stars) is necessary to deepen our  understanding of stellar evolution and important processes such as angular momentum transport \citep{doi:10.1146/annurev-astro-091918-104359}.

 In order to detect stellar pulsations, power spectra (squared absolute values of Fourier transforms of these recorded  lightcurves) are typically analyzed. The spectra typically show a sequence of peaks rising above a noisy background, with each peak corresponding to a globally oscillatory mode that may be characterized using three quantum numbers, radial order $n$, harmonic degree $\ell$, and azimuthal order $m$. The primary challenge is then to identify and label the peaks accurately. In red giants, this task can be difficult and time consuming if one performs it by visual inspection of the spectrum. Although semi-automated approaches have been developed \citep{period_spacing_2,Gehan2018,kallinger2019release}, the visual method remains common. Labeling allows to define appropriate assumptions in order to extract properties of modes such as the frequency, amplitude and width, generally using a fitting algorithm \citep{Vrard2018}. These parameters depend on the physical properties of the layers traversed by the modes, allowing us to infer the interior structure and rotation rates.
 
Current fitting methods to extract modes parameters in evolved stars e.g., \citep{bayesian_mcmc_2009,bayesian_mcmc_2011,bayesian_mcmc_2014} are relatively slow. This explains why only a small fraction of solar-like stars have been so far analyzed in details. To achieve a better understanding of stars in this era of ever-growing data, one needs efficient, much faster yet robust ways to measure features within the data that capture important physical information. 
 
Machine learning has emerged as a powerful tool with which to identify patterns in complex data sets. Here, we develop a machine learning algorithm that allows us to perform mode identification in one single extremely fast step, mimicking current fitting methods. In standard fitting methods, each star is individually analyzed (in some cases each mode); in contrast, machine learning enables the analysis of ensembles of stars at once, making it computationally efficient\footnote{Analysis of 1,000 stars takes $\sim$5s}. 


 In this article, we address the problem of measuring seismic parameters related to the structure of red giants using machine learning. 
Three seismic parameters are mainly involved - (a) large frequency separation ($\Delta \nu$), i.e., the average frequency spacing of $p$ - modes, (b) $\nu_{max}$, the $p$ - mode frequency at maximum power and (c) large-period separation ($\Delta \Pi$), i.e., the average period spacing of $g$ - modes. These parameters ($\Delta \nu$,$\nu_{max}$) are strongly correlated with mass and radius according to established scaling relations \citep{2012sse..book.....K,1991ApJ...368..599B,Mathur2012}. Along with this, the parameters ($\Delta \nu$, $\Delta \Pi$) separate the evolutionary stages of the star \citep{Mosser_2014} -  (i) subgiant, Hydrogen depletion phase in the core, (ii) red giant branch, the phase of H-burning in the shell (iii) red clump, the phase of He-burning in low-mass stars and (iv) secondary red clump, the phase of He-burning in high-mass stars.

\section{Results on real data} \label{Results on Real Data}

The success of machine learning entirely depends on the quality of the training data set. For the machine to detect a pattern and correctly predict a parameter in real data, we have modelled a synthetic data set that is realistic and that is able to account for typical variations in observations is required. We have incorporated the physics of structure, composition gradient, and rotation in red giants in our simulations using asymptotic theory of oscillations \citep{Garcia2019,2010aste.book.....A}. The detailed modelling of the synthetic data is given in the Appendix \ref{Methods and Techniques}. 
To construct the periodograms of \textit{Kepler} data\footnote{ Technical details of the procedure may be found in the Github repository. \url{https://github.com/OthmanB/LCconcat_kepler}}, we use the MAST data from which we extract the PDCSAP light-curves \citep{pdcsap1,pdssap2}, to which we fit a $6^{\mathrm{th}}$ order polynomial function in order to remove remaining trends in each  quarter. Following this, quarters are concatenated and data points that lie beyond 3.5$\sigma$ of the mean are discarded in order to filter out spurious data points. This post-processed light-curve is then used to compute the Lomb-Scargle periodogram \citep{Lomb1976,1982ApJ...263..835S} following the Rybicki Press algorithm \citep{1995PhRvL..74.1060R} prescription to calculate the Nyquist frequency.

\begin{figure*}[!t]
     \figurenum{1}
     \centering
     \includegraphics[width=170mm,scale=0.8]{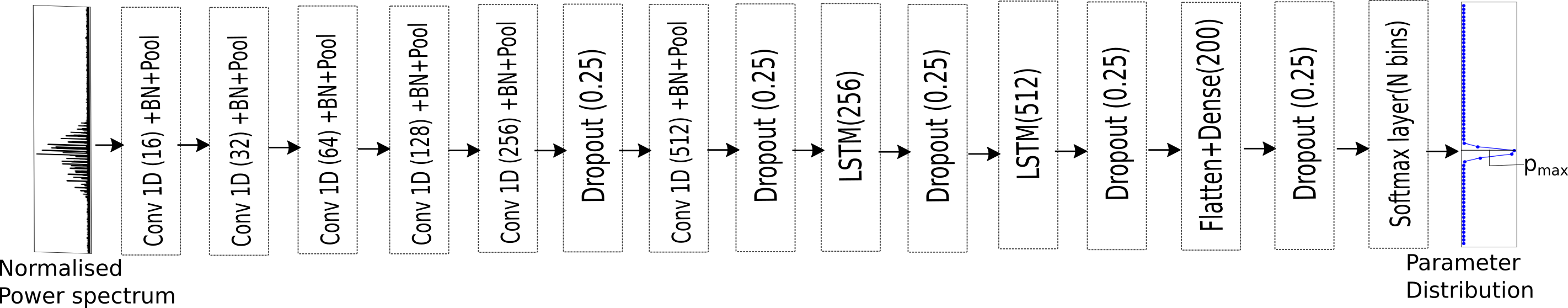}
     \caption{Architecture of neural network used in our machine learning algorithm. The network takes as input the (1D) normalized spectrum and outputs an approximate probability distribution of the subject parameter. $p_{max}$ in the inference of output shows the maximum probability (confidence) of the prediction. The core network is built using CNN, LSTM and Dense layers. Dropout layers with a fraction of 0.25 are used to prevent overfitting. The detailed network is presented in the Figure \ref{fig:ml_network_detail}}
    \label{fig:ml_network}        
        
\end{figure*}

We have trained neural networks to perform four different tasks. The first task being the detection of red giants and the other three measure the respective seismic parameters ($\Delta \nu$,$\nu_{max}$,$\Delta \Pi$). Although these networks are different, the architecture of these networks remain the same. We have built the network on a base of the Convolutional Neural Network. Figure \ref{fig:ml_network} shows the simplified architecture of the network, which takes a Normalized spectrum as an input and returns the inference of the seismic parameter. In the case of detection, the machine estimates the probability of Red-giant in the power spectrum. Additional details of the Machine learning model are provided in the Appendix \ref{Methods and Techniques}.

\begin{deluxetable*}{cccccc}
\tablecaption{Range of seismic parameters for the preparation of synthetic data.\label{tab:dataset_table}}
\tablewidth{0pt}
\tablehead{
\colhead{Parameter} & \colhead{Subgiant} & \colhead{Young red giant branch} & \colhead{Old red giant branch} & \colhead{Red clumps} \\
\colhead{} & \colhead{} & \colhead{(High-frequency red giants)} & \colhead{(Low-frequency red giants)} & \colhead{}
}
\startdata
Range of $\Delta \nu$ &  18-50 $\mu$Hz & 9-18 $\mu$Hz & 6-9 $\mu$Hz & 4.2-12 $\mu$Hz\\
\hline
Range of $\Delta \Pi$ & 60-200s  & 45-150s & 45-110s & 150-500s\\
\enddata
\tablecomments{This table shows the range of seismic parameters that were chosen to create different synthetic data sets. This range of parameters is chosen so as to cover the space of published results on \textit{Kepler} Data \cite{Mosser_2015,period_spacing_2,mixed_mode}. The detailed parameters used to create our simulations are shown in Table \ref{tab:dataset_table_extended}.}
\end{deluxetable*}

We first demonstrate that the machine can distinguish red-giant oscillation spectra from noise, allowing us to discover new red-giant stars from the \textit{Kepler} long cadence data set. We then validate machine inferences on \textit{Kepler} data and finally illustrate that the machine can identify the relationship between the seismic parameters.

\subsection{Machine-enabled detection of red giants}

\begin{figure*}[!t]
\figurenum{2}
\gridline{\fig{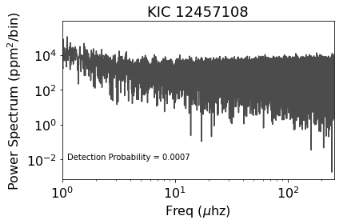}{0.46\textwidth}{(a)}
          \fig{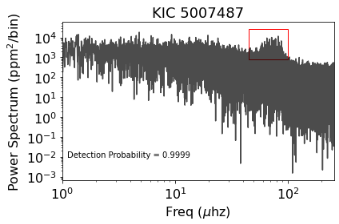}{0.46\textwidth}{(b)}
          }
\caption{(a) Oscillation spectrum of a noisy spectrum KIC 12457108. (b) Oscillation spectrum of a red-giant KIC 5007487.}
\label{fig:spec_noise_plots}
\end{figure*}

We first ensure that the machine is able to identify red-giant oscillation spectra. For this task, we have produced 500,000 synthetic data samples to train the neural network. The samples comprise an equal number of pure-noise and red-giant spectra, with a $\Delta \nu$ distribution of 1-18.7 $\mu$Hz and $\Delta \Pi$ ranging from 45-500s. Table \ref{tab:dataset_table_extended} shows the ranges of the seismic parameters in red-giant and noise simulations used to prepare the synthetic datasets. We then train the neural network to classify these samples. Figure \ref{fig:spec_noise_plots} shows examples of two stars,
one in which the spectra is dominated by noise (left) and another one in which the p-mode envelope is clearly visible (right). It shows that detection probability is low when the input is noise, whereas the network produces high probability for the p-mode envelope. If probability exceeds 0.5, we consider it to be a potential red giant. 

We apply this method to identify red giants from the ensemble of 151,000 \textit{Kepler} stars. Among these, 21,291 stars have been independently identified as red giants \citep{Hekker_2010,2013ApJ...765L..41S,2014ApJS..215...19P,2016ApJ...827...50M,Yu_2018,2018ApJS..239...32P,2019MNRAS.489.4641E,2020A&A...639A..63G,2020MNRAS.493.1388Y,2021A&A...648A.113B,Mosser_2015,period_spacing_2,mixed_mode,Hon2019}, out of which 17,527 stars are detected. Among other 130,288 stars not identified as pulsating red giants, 22,850 have been classified as positives. Thus, the algorithm shows 82.3\% and 17.5\% true and false-positive rates, respectively. 
Among these false positives, $\sim 50\%$ of stars are non-solar-like pulsators, such as  rapidly rotating, $\delta$-scuti, $\Gamma$-Doradus stars, etc. Thus, the false positive rate associated with mislabelling noise as red giant oscillations is $\sim$ 8\%. As part of future work, we will extend the preparation of synthetic datasets for non-solar-like pulsators to identify and categorize these stars. 

After visual inspection, we detect $\sim$25 new likely red giants using this method, shown in Table \ref{tab:table_redgiants_new}. We have explored various catalogues: \cite{Hekker_2010,2013ApJ...765L..41S,2014ApJS..215...19P,2016ApJ...827...50M,Yu_2018,2018ApJS..239...32P,2019MNRAS.489.4641E,2020A&A...639A..63G,2020MNRAS.493.1388Y,2021A&A...648A.113B,Mosser_2015,period_spacing_2,mixed_mode,Hon2019} to confirm that these 25 giant stars are new. We have provided the first measurements of $\Delta \nu$, $\Delta \Pi$, and $\nu_{max}$ for these stars in Table \ref{tab:table_redgiants_new}. Table \ref{tab:table_redgiants_2} provides the first measurements of $\Delta \nu$ and $\Delta \Pi$ for 195 stars, which were detected by \cite{Hon2019}. 

We thus establish that the machine can identify red giants. In the next subsection, we validate the seismic-parameter inferences by comparing them to other methods \citep{Mosser_2015, period_spacing_2, mixed_mode}. For this purpose, we show the results of machine inferences in 3,029 stars. We chose these stars as these are the only stars in \cite{Mosser_2015, period_spacing_2, mixed_mode} that fall into the parameter space of the training dataset shown in Table \ref{tab:dataset_table} and have reliable estimates of all the period-spacing parameters.

\subsection{Results of machine inferences} 
We have divided our synthetic training dataset into four subsets which indicate different evolutionary stages as given in Table \ref{tab:dataset_table}. For each stellar class given in Table \ref{tab:dataset_table}, we train the machine separately, leaving four different machines for each parameter, where the core networks are identical but the final layers change in accordance with the parameter and resolution required. Once trained, the neural network's performance is evaluated on a test data set, which comprises the unseen synthetic data. We show machine can infer $\Delta \nu$, $\nu_{max}$ and $\Delta \Pi$ successfully on this data set in Appendix \ref{Results_on_Synthetics}. 

To corroborate the deep-learning method, it is essential to verify the results from the neural network and compare them with estimates obtained using independent methods. We show that the trained machine can identify oscillation modes of \textit{Kepler} red giants. Though we have neural network outputs for 151,000 stars, for this analysis, we have selected 3,029 red giants from \cite{Mosser_2014,period_spacing_2} to allow a qualitative comparison between this work and other methods. Based on results from their analyses, we have categorized this 3,000-star sample into our four stellar classes:  \textit{subgiants} ($\Delta \nu$: 18-50$\mu$Hz), \textit{young red giant branch} ($\Delta \nu$: 9-18$\mu$Hz), \textit{old red giant branch} ($\Delta \nu$: 6-9$\mu$Hz, $\Delta \Pi$$<$150s), and \textit{red clumps} ($\Delta \nu$: 4.2-12$\mu$Hz, $\Delta \Pi$$>$150s). We show the performance of neural networks corresponding to each stellar class. 

\begin{figure*}[!t]
    \figurenum{3}
    \centering
    \includegraphics[width=160mm]{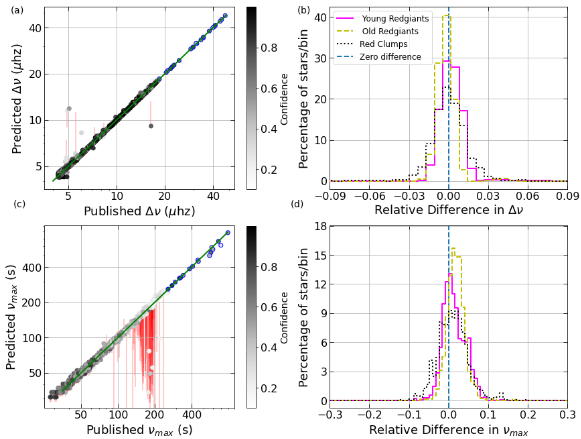}
    \caption{(a) Predicted value of $\Delta \nu$ at each value of published $\Delta \nu$ in all the detected stars by the neural network. The colour of each point in the plot denotes the confidence value of the prediction and the vertical red lines indicate 68\% confidence interval. Published values are taken from \cite{period_spacing_2}.(b) Distribution of relative differences in $\Delta \nu$ in various regimes of real data (legends). The relative difference is calculated with respect to the published value ((predicted $\Delta \nu$-published $\Delta \nu$)/published $\Delta \nu$). The red-dashed line tracks zero difference. More than 90\% of the predictions appear to lie well within 1.5\% of the published values.(c) Predicted value of $\nu_{max}$ at each value of published $\nu_{max}$ on all the detected stars by the neural network. The colour of each point in the plot denotes the confidence value of the prediction and the vertical red lines indicate 68\% confidence interval. Published values are taken from \cite{Yu_2018,Chaplin_2013}.(d) Distribution of relative differences in $\nu_{max}$ in various regimes of real data (legends). The relative difference is calculated with respect to the published value ((predicted $\nu_{max}$-published $\nu_{max}$)/published $\nu_{max}$). The blue-dashed line tracks zero difference. More than 90\% of the predictions appear to lie well within 5\% of the published values. The green line in (a,c) tracks \textit{predicted} parameter = \textit{published} parameter. Points with blue open circles show the predictions of \textit{subgiant} stars.}
    \label{fig:dp_dnu_all}        
\end{figure*}

Figure~\ref{fig:dp_dnu_all}(a) depicts neural-network predictions against the published values \citep{Mosser_2014,period_spacing_2} of $\Delta \nu$ in each stellar class. Figure \ref{fig:dp_dnu_all}(b) shows the distribution of relative differences between published values and corresponding neural network predictions of $\Delta \nu$. It indicates that $\Delta \nu$ is predicted well within 1.5\% of the published values for 90\% of the stars. In addition, they indicate that, apart from a few results on red-giant stars, the predictions are in agreement with published values. 

In Figure \ref{fig:dp_dnu_all}(c), $\nu_{max}$ predictions are plotted against the respective published $\nu_{max}$ values from \cite{Yu_2018} and \cite{Chaplin_2013}. Figure \ref{fig:dp_dnu_all}(d) graph the distributions of relative differences between $\nu_{max}$ predictions and published values in \textit{young red giant branch}, \textit{old red giant branch}, and \textit{red clumps} respectively. It indicates that $\nu_{max}$ is predicted well within 5\% of the published values for 90\% of stars. Therefore, these results along with $\Delta \nu$ predictions validate the neural network and also demonstrate that \textit{p} modes are encoded correctly in the  synthetic data set.

\begin{figure*}
\figurenum{4}
\gridline{\fig{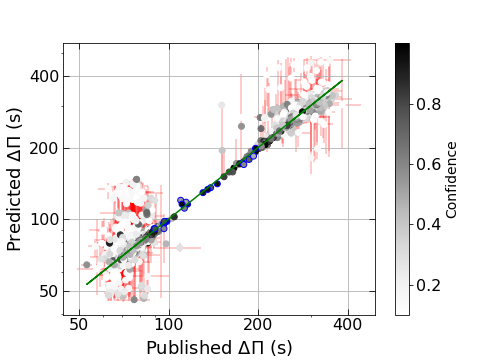}{0.46\textwidth}{(a)}
          \fig{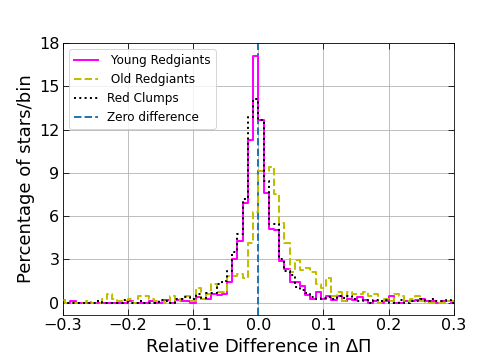}{0.46\textwidth}{(b)}
          }
     \centering
\caption{(a) Predicted value of $\Delta \Pi$ at each value of published $\Delta \Pi$ in all the detected stars by the neural network. The colour of each point in the plot denotes the confidence value of the prediction and the vertical red lines indicate 68\% confidence interval. The green line tracks \textit{predicted} $\Delta \Pi$ = \textit{published} $\Delta \Pi$. Points with blue open circles show the predictions of \textit{subgiant} stars. Published values are taken from \cite{period_spacing_2,Mosser_2014}. (b) Distribution of relative differences in $\Delta \Pi$ in various regimes of real data (legends). The relative difference is calculated with respect to the published value ((predicted $\Delta \Pi$-published $\Delta \Pi$)/published $\Delta \Pi$). The blue-dashed line tracks zero difference. More than 90\% of predictions appear to lie well within 5\% of the published values.}
\label{fig:dp_results_realdata}
\end{figure*}


Panel \ref{fig:dp_results_realdata}(a) capture $\Delta \Pi$ predictions against respective published $\Delta \Pi$ values in \textit{young red giant branch}, \textit{old red giant branch}, \textit{subgiants}, and \textit{red clumps} respectively. Panel \ref{fig:dp_results_realdata}(b) graph the distributions of relative differences between predicted and published $\Delta \Pi$ in each stellar class. The distributions of relative differences show that, for 90\% of the stars, the neural network recovers $\Delta \Pi$ to within 7\% of published values \citep{Mosser_2014,period_spacing_2}. These results also indicate that the machine works very well in predicting $\Delta \Pi$ in sub giants, young red giant branch and red clumps. Figure~\ref{fig:dp_results_realdata}(d) shows that it is not quite as successful when applied to old red giants as well as to other evolutionary stages. In the old red-giant branch, 90\% of the predictions are within 12\% of the published results. This relatively poor performance may be ascribed to the lower mixed-mode coupling strengths for old red giants \citep{mixed_mode}, contributing to the comparatively diminished performance on synthetic data (Figure \ref{fig:dp_results_syn}) and the training data requiring a larger variety of templates. 

In low-amplitude dipolar-mode stars e.g. \citep{depressed_modes}, the neural network identifies these as solar-like due to the p-mode hump. Additionally, the network's inference of $\Delta \Pi$ in these stars will not be reliable, as they are not modelled in the synthetic dataset.

Therefore, the method is highly accurate in \textit{young red giant branch}, \textit{subgiants}, and \textit{red clumps}. In these stellar classes, 90\% of the $\Delta \Pi$ predictions agree with published values to within 7\%. This method is moderately accurate for the \textit{old red giant branch} ($\Delta \nu<9\mu$Hz and $\Delta \Pi<150$s).  In this stellar class, 90\% of the $\Delta \Pi$ predictions agree with published values to within 12\%. These results also prove that mixed modes are encoded correctly in the synthetic data set.

\subsection{Evolution of stellar and core density in a red-giant}

\begin{figure}[!t]
     \figurenum{5}
     \centering
     \includegraphics[width=78mm]{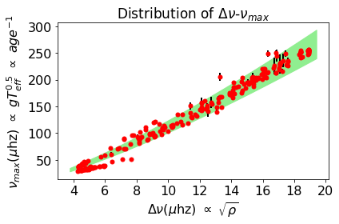}
     \caption{The distribution of $\Delta \nu$-$\nu_{max}$ for the stars listed in Tables \ref{tab:table_redgiants_new} and \ref{tab:table_redgiants_2}. The red points correspond to ($\Delta \nu$, $\nu_{max}$) and the green band maps the relation given in \cite{Stello_2009}. The black lines associated with each point mark the 1-$\sigma$ interval. In most cases, the 1-$\sigma$ intervals are smaller than the sizes of the plotted points, and hence not visible to the naked eye. These parameters depend on stellar density ($\rho$) and age of the star. This plot indicates stellar density decreases as the star evolves.}
    \label{fig:dnu_numax_dist}      
\end{figure}

\begin{figure*}[!t]
    \figurenum{6}
    \centering
    \includegraphics[width=140mm]{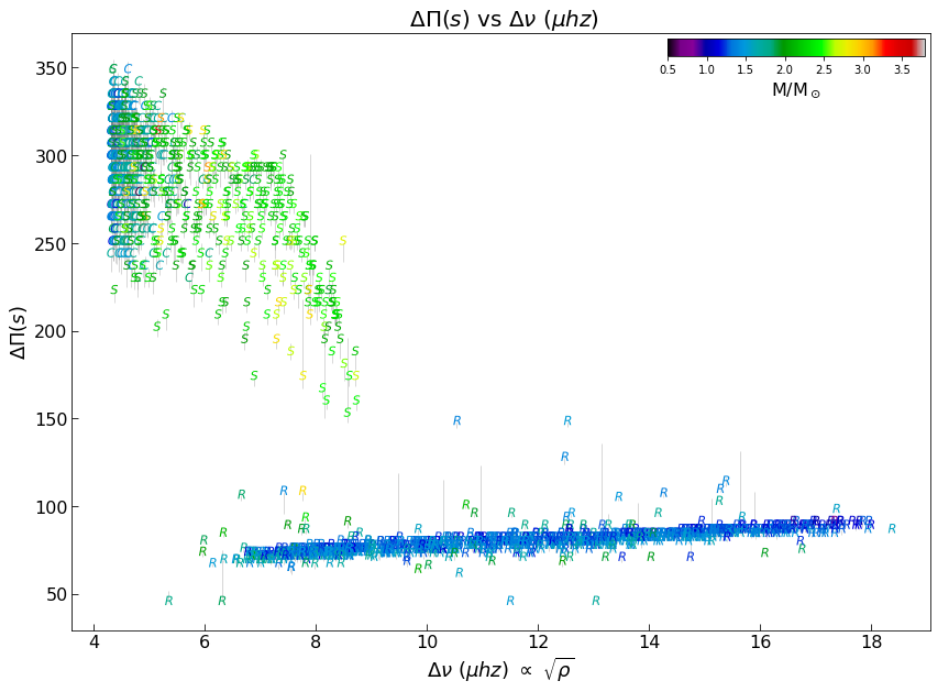}
    \caption{Machine $\Delta \Pi$ vs machine $\Delta \nu$ predictions. The colour of each point captures the ratios of stellar to solar masses (stellar masses are taken from \cite{period_spacing_2}). The letters R,S,C represent red giants, secondary red clumps and red clump stars respectively. The grey lines demarcate 68\% confidence intervals. All these predictions are highly confident. The plot indicates that there exists a nearly linear relationship between $\Delta \Pi$ and $\Delta \nu$ in red giants. $\Delta \nu$ depends on stellar density ($\rho$) of the star and $\Delta \Pi$ depends on core sizes of stars \citep{core_size_DP}. This plot indicates core size decreases as the stellar density decreases in red giants.}
    \label{fig:dp_dnu_rg}        
\end{figure*}

Figure~\ref{fig:dnu_numax_dist} shows the $\Delta \nu$-$\nu_{max}$ distribution of stars given in Tables \ref{tab:table_redgiants_new} and \ref{tab:table_redgiants_2}.  It shows that most stars follow the relation given in \cite{Stello_2009}. However, at very low $\Delta\nu$, we note that several stars deviate by more than $1\sigma$ from the general trend, indicating a possible break of the scaling relation for most evolved stars. This deviation has also been observed for evolved stars in \cite{Huber_2011}, where they have observed a different $\Delta \nu - \nu_{max}$ relation for stars with $\nu_{max}<100\mu$Hz. This deviation is not an artifact of periodogram construction, as we have independently verified the predictions on periodograms from the lightkurve software \citep{2018ascl.soft12013L}. These parameters depend on stellar density ($\rho$) and age respectively \citep{2012sse..book.....K}. The plot shows that, as the star evolves, the stellar density decreases.

Figure~\ref{fig:dp_dnu_rg} presents the $\Delta \Pi$-$\Delta \nu$ relation in red giants and red clumps. It shows that $\Delta \Pi$ and $\Delta \nu$ are approximately linearly related in red giants. $\Delta\Pi$ depends on the core size of the star \citep{core_size_DP} and this plot shows that the core contracts as stellar density decreases with progressing stages of evolution. Even though this result is well established \citep{Mosser_2014}, it is important for the following reasons:
\begin{itemize}
    \item It serves as a validation test for the neural network and indicates high-quality synthetic data.
    \item  The training data sets that were constructed do not have a built-in correlation between $\Delta \Pi$ and $\Delta \nu$. However, when applied to real data, the neural network finds a strong correlation between these two parameters ($\Delta \Pi$, $\Delta \nu$). This demonstrates that the machine is able to find the true correlation between these parameters in quick computational time. 
\end{itemize}

\subsection{Rare Systems}
\begin{figure*}[!t]
\figurenum{7}
\gridline{\fig{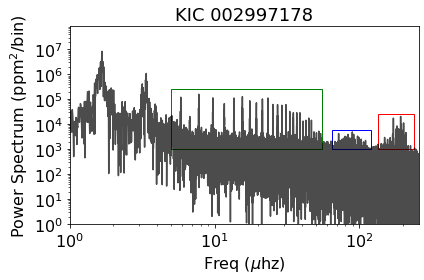}{0.46\textwidth}{(a)}
          \fig{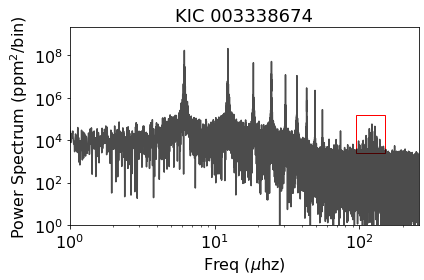}{0.46\textwidth}{(b)}
          }
\caption{(a) Power spectrum of KIC 2997178. The red box in the power spectrum shows the red-giant component. The blue and green boxes in the plot highlights other two features of the power spectrum.(b) Power spectrum of the binary KIC 3338674 where the red-box in the power spectrum highlights the red-giant component.}
\label{fig:special_systems_a}
\end{figure*}

As the four stellar classes given in the Table \ref{tab:dataset_table} have stellar oscillations in different frequency ranges, the trained networks survey different parts of the power spectrum. Therefore, this method can identify red giants in binaries as well. Table \ref{tab:table_redgiants_new} has four binary candidates KIC 2997178, KIC 3338674, KIC 2018906 and KIC 1295546, which are shown in Figures \ref{fig:special_systems_a} and \ref{fig:special_systems_b}. Figure \ref{fig:special_systems_a}(a) shows the power spectrum of KIC 2997178 highlighting three different parts of the power spectrum. The possible explanations for the three features of this power spectrum are: 
\begin{itemize}
    \item A red giant (marked in red) eclipsing another other star (marked in blue). The green box in this Figure \ref{fig:special_systems_a}(a) highlights the signal of this eclipse.
    \item An eclipsing binary candidate (green, blue) with a red giant in the background (red).
    \item A possible triplet which has a red giant.
\end{itemize}

\section{Conclusion}

We present an efficient machine-learning algorithm that learns the marginal distributions of global seismic parameters $\Delta \nu$, $\nu_{max}$, and $\Delta \Pi$. We create a library of synthetic data to train the machine and demonstrate its performance on oscillation spectra of stars in different stages of evolution. The network is calibrated and able to identify \textit{p}- and mixed-mode patterns on synthetic data, since it accurately finds $\Delta \nu$, $\nu_{max}$, and $\Delta \Pi$.

The machine can successfully discern red giants from noise on \textit{Kepler} data. In addition to the 17527 known stars that the machine has also identified, we have identified $\sim 25$ new red giants analysing 151,000 \textit{Kepler} long-cadence stars in a few minutes of computation. We have provided the first period-spacing measurements of the $25$ new red giants and those of 195 red giants previously identified by \cite{Hon2019}. Although not being trained explicitly to, the present method is still capable of detecting red giants in binaries. We have detected four new binaries with red-giant counterparts.

We validate the machine inferences using estimates from other methods by selecting $\sim$3,000 test stars from  \cite{Mosser_2014,period_spacing_2}, which are spread over a range of evolutionary stages. We observe that 90\% of the predictions agree with published values to within 1.5\% for $\Delta \nu$, 8\% for $\Delta \Pi$ and 5\% for $\nu_{max}$. 
Despite the training data containing no artificially introduced correlations among the seismic parameters, the machine has independently identified a linear relationship between $\Delta \Pi$ and $\Delta \nu$ in the observations, validating both the neural network and the synthetic spectra. 

In future work, we will improve the machine learning model and synthetic data to improve estimates of $\Delta \Pi$, and expand the parameter space to include stars with $\Delta \nu <4.2 \mu Hz$. We will investigate the undetected stars and special systems detected by the machine. We will optimize the training time of our machine learning model using multi-node setup. Also, the model  will be expanded to infer all global seismic parameters, such as the coupling constant, core and envelope rotation rates, and inclination angle, by combining this method with Monte-Carlo-based techniques \citep{bayesian_mcmc_2009,bayesian_mcmc_2011,bayesian_mcmc_2014}.

The neural network can study $\sim$1,000 stars in under $\sim$5 seconds, enabling ensemble asteroseismology on vast data sets. As the machine is completely trained on synthetic data, we can extend this to data from other missions with small changes (such as including mission-specific systematics) to the simulations. Future missions such as \textit{PLATO} are expected to observe a million light curves, which might consist $>100,000$ solar-like stars. The network here can analyze and extract the physics of these stars within 10 minutes, and has the potential to transform asteroseismology. 

\textbf{Acknowledgment:} S.D. acknowledges SERB, DST, Government of India, CII and Intel Technology India Pvt. Ltd. for the Prime minister's fellowship and facilitating research. All the computations are performed on Intel\textsuperscript{\textregistered} Xeon\textsuperscript{\textregistered} Platinum 8280 CPU. We thank Dhiraj D. Kalamkar, Intel Technology India Pvt Ltd for the suggestions, which helped to optimize the neural network training. This paper includes data collected by the \textit{Kepler} mission and obtained from the MAST data archive at the Space Telescope Science Institute (STScI). Funding for the \textit{Kepler} mission is provided by the NASA Science Mission Directorate. STScI is operated by the Association of Universities for Research in Astronomy, Inc., under NASA contract NAS 5–26555. This research made use of Lightkurve, a Python package for \textit{Kepler} and TESS data analysis (Lightkurve Collaboration, 2018). We thank Tim Bedding and the anonymous referee for providing constructive comments, which helped improve the quality of the paper.

\bibliography{scibib}{}
\bibliographystyle{aasjournal}

\appendix
\counterwithin{figure}{section}
\counterwithin{table}{section}
\twocolumngrid

\section{Methods and Techniques}\label{Methods and Techniques}
\subsection{Simulated spectra as a training data set}

We build data sets using a simulator available\footnote{The version used for this paper is available in the {\it Siddarth2021} branch.} at \url{https://github.com/OthmanB/Spectra-Simulator-C} that can generate synthetic spectra over a large range of parameters. The software incorporates the physics of structure, composition gradient, and rotation in red giants using the asymptotic theory of stellar oscillations \cite{Garcia2019,2010aste.book.....A}. The simulator takes a random global seismic parameter set over a range specified by the user and generates a spectrum.  For a specific set of parameters, different noise realisations are generated in order to train the machine to discriminate features from noise. In this section, we describe the asymptotic theory and preparation of synthetic data sets to train the machine 

\subsubsection{Frequencies of {\it p} and mixed modes} \label{sec:freq_p_mixed_modes}

Global stellar oscillations are predominantly due to standing waves of two kinds, one where pressure is the restoring force ({\it p} modes) and the other where buoyancy is the restoring force ({\it g} modes). While {\it p} modes can travel all through the interior, pure {\it g} modes are trapped in the deep radiative zone and have surface amplitudes far too small to be observed. This is due to the fact that solar-like stars have a thick outer convective zone in which {\it g}-mode oscillations are evanescent. Ever since early theoretical work on this topic \cite{1989nos..book.....U}, it is known that when the physical distance between the cavities of {\it p} modes and {\it g} modes become small enough or overlap, the modes may significantly interact to form so-called \textit{mixed modes}. Unlike {\it p} modes that mostly probe outer convective layers, mixed modes provide a unique window into deep internal structure. Observational asteroseismology has revealed that prevalent conditions in red giants allow for the existence of mixed modes \cite{Bedding2010, 2011Sci...332..205B}.

For unresolved disk photometry,  degrees higher than $\ell>3$  modes cannot usually be observed due to geometrical cancellation effects that limit their apparent amplitude (see Section \ref{Relative heights and widths of different modes}).

In the case of a spherically symmetric, non-rotating star, $m$ components are degenerate and frequencies only depend on the degree and radial order. {\it p}-mode frequencies are then expected to approximately follow an asymptotic regime \cite{Tassoul1980}. In second-order asymptotic theory, the frequencies of \textit{p} modes of radial order \textit{n} and azimuthal order $\ell$ are given by \cite{Mosser_2010,Mosser_2012}
\begin{equation}
    \frac{\nu_{n,\ell}}{\Delta \nu} = n + \frac{\ell}{2}+ \epsilon(\Delta \nu) - d_{0\ell}(\Delta \nu) + \frac{\alpha_{\ell}}{2}\left(n-\frac{\nu_{max}}{\Delta \nu}\right)^2,
    \label{eq:nu_p}
\end{equation}

where $\Delta \nu$ is the \textit{large-frequency separation}, which gives the mean-frequency separation between two successive radial modes, $\epsilon(\Delta \nu)$ is the \textit{offset parameter}, $d_{0\ell}$ the \textit{small-frequency separation}, and $\alpha_{\ell}$ the degree-dependent gradient $\alpha_{\ell} = \left(d\log \Delta \nu/dn\right)_{\ell}$.

\begin{figure}
     \centering
     \includegraphics[width=85mm]{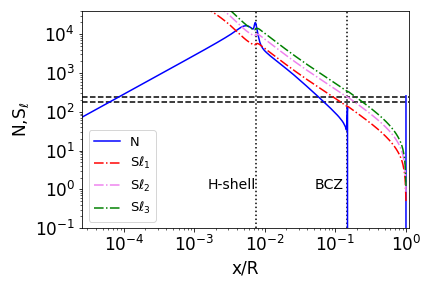}
     \caption{Wave propagation diagram in a 1.3M$_\odot$ star at 4.51 Gyr, in a young red-giant phase}; plotted are Lamb and Brunt-Väisälä frequencies as a function of the fractional radius. Lamb frequencies determine the \textit{p}-mode cavities of the respective $\ell$ modes and Brunt-Väisälä frequencies set the \textit{g}-mode cavity. The horizontal dashed lines denote the ranges of observable frequencies for this red-giant star. The vertical dotted lines demarcate the Hydrogen shell and base of convection zone.
    \label{fig:propagation_diagram}
\end{figure}

The term $\nu_{max}$ in Equation \ref{eq:nu_p} refers to the frequency corresponding to maximum amplitude. Observations \cite{Stello_2009} and scaling relations \cite{1986ApJ...306L..37U} have demonstrated a strong interdependent relation between $\Delta \nu$ and $\nu_{max}$. For our simulations, we choose $\nu_{max}$ based on $\Delta \nu$ from the relation given in \cite{Stello_2009}, with  10\% deviation, as follows
\begin{equation}
    \nu_{max} = (\Delta \nu / 0.263)^{1/0.772} \pm 0.1 (\Delta \nu / 0.263)^{1/0.772}.
    \label{eq:eq_dnu_numax}
\end{equation}
For solar-like stars, mixed modes start to become visible in the power spectra when a star reaches the end of the main-sequence. There is an increase in the density gradient in the core, which causes the Brunt-Väisälä frequency ($N$) to rise. As illustrated in Figure \ref{fig:propagation_diagram}, one consequence is that the acoustic and buoyancy cavities become closer to each other. As a result, the coupling between the interior \textit{g} modes and \textit{p} modes grows stronger. The strength of coupling is indirectly proportional to the physical distance between these cavities \cite{mixed_mode}. Mixed modes exhibit characteristics of both types of oscillations at the same eigenfrequency. This mode is oscillatory in the radiative core and the acoustic envelope but evanescent in the region that connects the cavities. 

Largely, the observed mixed modes are dipole modes as dipole \textit{p} modes travel deeper into the star and hence have higher coupling strength with dipole \textit{g} modes compared to $\ell=2,3$ modes. Figure~\ref{fig:propagation_diagram} shows that dipole \textit{p}-mode cavity probe deeper than $\ell=2$ and $\ell=3$ \textit{p} mode cavities. Even though there are observed cases of $\ell=2$ mixed modes \cite{Benomar_2013}, they are exceptions and have very low coupling strength. Therefore, in our simulation, we consider $\ell=2$ and $\ell=3$ as pure \textit{p} modes and dipole modes as mixed modes. The coupling phenomena between {\it p} and {\it g} modes was theorised by \cite{1989nos..book.....U} and formally expressed in the case of evolved solar-like stars using asymptotic relations for {\it p} and {\it g} modes by \cite{Mosser_2012}. They derived the following implicit equation that gives the solutions of dipole mixed-mode frequencies,
\begin{equation}
    \tan \pi \frac{\nu-\nu_p}{\Delta\nu} = q \tan \frac{\pi}{\Delta \Pi} \left(\frac{1}{\nu}-\frac{1}{\nu_g}\right),\hspace{0.2cm} \label{eq:mixedmodes}
\end{equation}
where $q$ is the coupling factor between {\it p} and {\it g} modes, $\nu_p$ is the frequency of a pure \textit{p} mode, $\nu_g$ is the frequency of a \textit{g} mode, and $\Delta \Pi$ is the \textit{period spacing} that depends on the integral of the Brunt-Väisälä frequency and is therefore sensitive to the deep interior structure of solar-like stars. It defines the period separation between successive pure \textit{g} modes, asymptotically regularly spaced in period, 

\begin{equation}
    \frac{1}{\nu_g} = \left(-n_{g} + \epsilon_{g}\right)\Delta\Pi
    \label{eq:nu_g}
\end{equation}

where $n_g<0$ is the radial order of the pure \textit{g} mode and $\epsilon_g$ is the \textit{offset parameter}. When solving equation \ref{eq:mixedmodes}, the second-order asymptotic relation shown in equation \eqref{eq:nu_p} is used for {\it p}-mode frequencies as well as the equation \ref{eq:nu_g} for the {\it g} modes frequencies.

Due to the difference in gradients of the Brunt-Väisälä and Lamb frequencies, the coupling factor is expected to vary. However, in the simulations, it is assumed that these variations remain small within the range of observed frequencies so that $q$ is set to be constant. This commonly invoked assumption has been shown to be accurate in past studies \cite{Mosser_2015}.

\subsubsection{Effect of Rotation}
Evolved solar-like stars are known to show significant radial differential rotation \cite{six_stars}. The core-to-envelope rotation contrast can range up to a few tens, and its imprint on the pulsation frequencies must be taken into account in realistic simulations.

While the asymptotic relations described in Section \ref{sec:freq_p_mixed_modes} remain valid, rotation breaks the spherical symmetry of the star and lifts the degeneracy in $m$. The Sun shows radial and latitudinal differential rotation \citep[e.g.,][]{1998ApJ...505..390S}, with a mean sidereal rotation rate of $\sim$435 nHz (the $a_1$ coefficient). Slow rotation, as in the Sun, may be treated as a small perturbation to the non-rotating case. Each mode of degree $\ell$ splits into $2\ell+1$ azimuthal components, with $m\in[-\ell, \ell]$. The mode $\nu_{n,\ell,m}$ is given by $\nu_{n,\ell,m}=\nu_{n,\ell}+\delta\nu_{n,\ell,m}$, with $\delta\nu_{n,\ell,m}$ being the {\it rotational splitting}. Considering that the radial differential rotation dominates relative to latitudinal differential rotation, it may be expressed as a weighted average of the rotational profile,
\begin{equation}
    \delta\nu_{n,\ell,m} = m \int^{R}_{0} K_{n,l}(r) \Omega(r) dr,
\end{equation}
where $K_{n,l}$ is the rotation kernel, defining the sensitivity of the modes as a function of radial position $r \in [0,R]$ within the star.

Because {\it p} modes are mostly sensitive to average rotation within the stellar interior \cite[]{Benomar2015}, the dependence of $\delta\nu_{n,\ell,m}$  on $(n,\ell)$ is weak within the observed frequency range of a solar-like star \cite{Lund2014b}. In fact, it does not exceed a few percent even in the presence of radial differential rotation of a factor two between the convective zone and the radiative zone. Note that in the Sun, this differential rotation is of $\simeq 30\%$, while on other solar-like stars, it is below a factor two \cite{Benomar2015,Nielsen2017}. Such upper limit factor leads to splitting variations of the same other as that achieved in the 1$\sigma$ uncertainties of the best seismic observations from \textit{Kepler}. It is accurate to express the rotational splitting of {\it p} modes as

\begin{equation}
    \nu_{n,\ell,m}=\nu_{n,\ell}-m\nu_s,
    \label{eq:splitting_pmodes}
\end{equation}
where $\nu_s=\Omega/2\pi$ is the integral term of equation \ref{eq:splitting_pmodes}, a function only of the average internal rotation $\Omega$ rate. The rotational kernels of {\it p} modes are highly sensitive to outer layers of stars. In main-sequence stars, approximately $60\%$ of the average rotation rate $\Omega$ comes from the contribution of the envelope rotation. For red giants, which have a much larger envelope than main-sequence stars, this contribution exceeds $80\%$, e.g., Figure \ref{fig:propagation_diagram}. And it is common to consider that $\nu_s$ essentially measures the rotation in the envelope, so that $\nu_s\simeq\Omega_{env}/2\pi$ \citep{goupil_env_rotation}.

The $\ell = 1$ modes and $\ell=2,3$ modes are affected differently by the rotation in red giants. $\ell=2,3$ modes are considered to be pure \textit{p} modes so that frequencies of split components are considered to follow equation \ref{eq:splitting_pmodes}.

The mixed modes are primarily present in the $\ell=1$ oscillations and these are influenced by both the core (\textit{g} modes) and the envelope (\textit{p} modes). Due to the mode mixing, Kernels $K_{n,\ell=1}(r)$ differ significantly from one mode to another. Some mixed modes indeed are weakly sensitive to the core while others show strong sensitivity. However, \cite{Goupil2013} has shown that a two-zone model of rotation can well account for $\ell=1$ splittings observed in red giants and early subgiants. Under that assumption, they also demonstrated that rotational splitting is a linear function of the ratio between the kinetic energy of modes in the g modes cavity and the total kinetic energy of modes, denoted $\zeta(\nu)$,
 \begin{equation}
    \mathrm{\delta\nu_{rot}} = -\frac{1}{2}\frac{\Omega_{core}}{2\pi}\zeta(\nu) + \frac{\Omega_{env}}{2\pi} (1-\zeta(\nu)).
\end{equation}

Furthermore, it was found \cite{Deheuvels2015} that the $\zeta(\nu)$ function is well approximated by, 
\begin{equation}
    \zeta(\nu) = \left[1+\frac{1}{q}\frac{\nu^2\Delta \Pi}{q\hspace{0.05cm}\Delta \nu } \frac{\cos^2\pi \frac{1}{\Delta \Pi} \left(\frac{1}{\nu} - \frac{1}{\nu_g}\right)}{\cos^2\pi \frac{\nu-\nu_p}{\Delta \nu}}\right]^{-1},
\end{equation}
which is defined by the same quantities as equation \ref{eq:mixedmodes}.
When $\zeta(\nu)$ is close to one, the mode is mainly trapped in the g-mode cavity (and thus is more sensitive to the stellar core). A value of $\zeta(\nu)$ close to 0 correspond to a mode essentially trapped in the p-mode cavity.

This expression is broadly used to determine the rotational splittings of red giants by various authors such as \cite{Mosser2018}, but it is a crucial relation that also describes the observed period spacing \citep{Gehan2018}, the amplitudes and the width variations as functions of the mode frequency in evolved stars (see Section \ref{Relative heights and widths of different modes}).

\subsubsection{Relative heights and widths of different modes}
\label{Relative heights and widths of different modes}

\begin{figure*}
\gridline{\fig{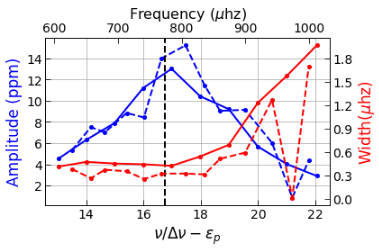}{0.46\textwidth}{(a)}
          \fig{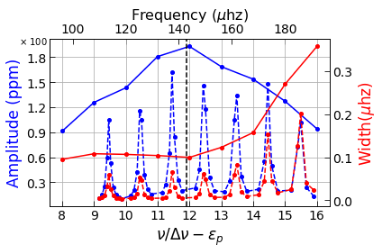}{0.46\textwidth}{(b)}
          }
\caption{This figure demonstrates the conversion of the observed spectral template (KIC 12508433) into a synthetic spectral template. (a) Amplitude and width profiles of KIC 12508433. (b) Amplitude and width profiles of a synthetic star ($\Delta \nu= 12\mu$Hz, $\Delta \Pi= 80$s, $q=0.1$, $\Omega_{core}/2\pi = 1\mu $Hz, $\Omega_{env}/2\pi = 0.05 \mu$Hz, $\iota=54.4 ^{\circ}$) that is based on the template of KIC 12508433. The solid (dashed) lines in both figures correspond to $\ell=0$ ($\ell=1$) modes. The blue (red) lines in both figures correspond to amplitude (width) profiles. }
\label{fig:template_real_synthetic_star}
\end{figure*}

Each mode in the power spectrum is modeled using a Lorentzian function centered around frequency $\nu(n,\ell,m)$ with height $H(n,\ell,m)$ and width $\Gamma(n,\ell,m)$. The linewidths $\Gamma(n,\ell)$ of \textit{p} modes in the power spectrum increase with frequency \citep{width_constant}. The excitation and damping of modes is not well understood, mostly due to non-adiabatic effects near the stellar surface, which are difficult to model. Yet our simulations need to have realistic mode heights and widths. To circumvent this issue, we use templates obtained by fitting real stars and rescale them following an adapted version of the method from \cite{Kamiaka2018}. 
In this technique, the heights, widths and average splittings of p modes are rescaled. For l=0,2,3 p modes, the procedure here is similar. However because individual pulsation frequencies of the template are different than those of the simulated star (and given by equation \ref{eq:nu_p}), it is important to correct for differences in $\nu_{max}$ and $\Delta \nu$. Linear interpolation is performed to stretch and recenter the template height and width at the frequencies of the simulated star. This allows us to obtain profiles for $\ell=0,2,3$ p modes that are identical to that of the template but recentered at the $\nu_{max}$ of the simulated star\footnote{Technical details of the procedure may be found in the Github repository. \url{https://github.com/OthmanB/Spectra-Simulator-C/bump_DP.cpp}.}.

As revealed by \cite{Benomar_2013,Grosjean2015}, mixed modes show complex amplitude and width variations with frequency. Motivated by this observational evidence, \cite{Benomar2014} found that the mode inertia ratios between $\ell=1$ and $\ell=0$ modes are expressed as
\begin{equation}
    \frac{I_1}{I_0} = \sqrt{\frac{A_0}{A_1}} \frac{\Gamma_0}{\Gamma_1},
    \label{eq:inertia_ratio1}
\end{equation}
where $A_0\propto \sqrt{\pi H_0 \Gamma_0}$, $A_1\propto\sqrt{\pi H_1 \Gamma_1}$ denotes the amplitudes of $\ell=0$ and $\ell=1$ modes and $\Gamma_0$, $\Gamma_1$ are the mode widths.
\cite{Grosjean2015b} found that, under the assumption of equipartition of energy between modes, and accounting for damping and excitation assuming no radiative pressure, the product of inertia and width is conserved between $\ell=0$ and $\ell=1$ modes, 
\begin{equation}
    I_1 \Gamma_1 = I_0 \Gamma_0.
    \label{eq:inertia_ratio2}
\end{equation} 
Furthermore, \cite{Mosser_2015} found that the inertia ratio may be expressed as a function of $\zeta(\nu)$,
\begin{equation}
    \frac{I_1}{I_0} = \frac{1}{1-\zeta(\nu)}.
    \label{eq:inertia_ratio3}
\end{equation}

Equations \ref{eq:inertia_ratio1}, \ref{eq:inertia_ratio2}, \ref{eq:inertia_ratio3} may be used to derive the amplitudes and widths of the mixed modes,
\begin{equation}
    \Gamma_1 (\nu) = \Gamma_0 (1-\zeta(\nu));\hspace{0.2cm}A_1^2 (\nu) = A_0^2 (1-\zeta(\nu)).
\end{equation}

Due to the assumptions made for reaching this expression, it may be accurate only for less evolved stars, i.e., red giants and subgiants. However, as shown in Section \ref{Results on Real Data}, it remains accurate enough for red clump stars as well, ensuring that the current machine-learning approach does not introduce biased results on fundamental quantities defining the mixed-mode frequencies.

Figure \ref{fig:template_real_synthetic_star} illustrates an example of this procedure of producing a template for synthetic star. It exhibits an amplitude and width profile of KIC 12508433, and shows a template prepared for the synthetic star. The blue (red) solid lines represent amplitude (width) of the $\ell=0$ modes. It can be observed that amplitudes of $\ell=0$ modes are nearly symmetric with respect to $\nu_{max}$ in both cases. Also, the amplitudes and widths of the template follow the same trend as the original star, establishing the method of conversion described in this subsection. 

\subsubsection{Effect of mode visibility and of stellar inclination} 

When observing oscillations of distant stars at low spatial resolution, the visibility of mode $f_{n,\ell,m}=AY_\ell^m(\theta,\phi)$ is given by 
\begin{equation}
    a_{n,\ell,m}=r_{\ell,m}(\iota)V(\ell) A,
\end{equation}
where $V(\ell)$ is the mode visibility, $r_{\ell,m}(\iota)$ the relative amplitude of the mode, which depends on the inclination angle $\iota$. The visibility function depends on the limb-darkening function (star type) and the measurement technique used.

The visibility function $V(\ell)$ decreases with increasing degree $\ell$. Therefore, we dominantly observe only $\ell=0,1$ and $2$ modes in the asteroseismic data, as the amplitude decreases for other degree modes; the $\ell=1$ mode has higher visibility than $\ell=0$. We rarely observe $\ell=3$ modes. From studies on various red giants \cite{Mosser2012a}, it is assumed that $V(0) \simeq 1$, $V(1) \in [1.2,1.75]$, $V(2) \in [0.2,0.8]$ and $V(3) \in [0,0.1]$.

The relative amplitude is given by the following equation:
\begin{equation}
    r_{\ell,m}^2\left(\iota\right) = \frac{(\ell-|m|)!}{(\ell+|m|)!}\left[P_\ell^{|m|}(\cos \iota)\right]^2,
\end{equation}
where $P_\ell^{|m|}$ is the associated Legendre polynomial. To maintain an isotropic distribution of stellar-axis inclinations, the prior for the angle is taken to be $P(\iota) \propto \sin(\iota)$.

\subsubsection{Noise Model}

The background noise model comprises a combination of white noise and a Harvey-like profile. At high frequencies, noise is dominated by white noise (photon noise), which is independent of frequency. At low frequencies, noise is generated by surface convection (granulation), described by the Harvey profile \cite{harvey_profile}. While there are models considering additional facular signatures \cite{Karoff_2013}, these are not the dominant features \cite{Karoff_2012} and hence are ignored here. Equation \ref{eq_noise_model} contains the background noise model $B(\nu)$:
\begin{equation}
    B(\nu) = \frac{H}{1+(\tau\nu)^p} + N_0,
\label{eq_noise_model}
\end{equation}
where H is the characteristic granulation amplitude, $\tau$ is the characteristic timescale of granulation, $p$ is the characteristic power law, and $N_0$ is the white noise level. 

It has been observed that the granulation amplitude and timescale vary with $\nu_{max}$ \cite{Kallinger_2010,Mathur_2011,Chaplin_2011}. Considering this, we model the granulation amplitude and timescale as 
\begin{equation}
     H = A_{g} \nu_{max}^{B_{g}} + C_{g},
     \,\,\,\,\,
     \tau = A_{\tau} \nu_{max}^{B_\tau} + C_\tau,
     \label{gran_time}
\end{equation} 

where the triplet ($A_{g},B_{g},C_{g}$) has free parameters that modify granulation amplitude and  ($A_\tau,B_\tau,C_\tau$) modify the granulation timescale. We have created a wide variety of noise profiles by taking a range of values for these parameters, as given in Table \ref{tab:dataset_table}. 

\subsubsection{Description of the Data sets}
\label{Description of Data set}

We generate 3 million random synthetic stellar spectra\footnote{ Requiring $\sim$ 20,000 core hours ($\sim$ 125hrs $\times$ 160 CPUs).} spanning the range of seismic parameters described in Table \ref{tab:dataset_table}. The spectra in the data set possess a variety of different features apart from the primary set of seismic parameters: (a) variable number of peaks, (b) various height profiles, (c) variable resolution and (d) uniform prior in parameter space.
 
In step (a), stellar spectra in the data set have different numbers of modes since real data show differing numbers of peaks. This step is important as parameter prediction must be robust to changes in the magnitude of the star and signal-to-noise ratio. As the magnitude of the star and SNR rise, the number of observed peaks increases.

For step (b), we generate the data set based on nine amplitude profiles of different stars. Each spectrum randomly selects an amplitude profile among these stars and creates a template according to the method described in Section \ref{Relative heights and widths of different modes}. As our parameter set is a function of the peak positions, this step ensures that parameter prediction is independent of height. To create this data set, we have selected the following profiles: KIC 10147635, KIC 11414712, KIC 12508433, KIC 6144777, KIC 8026226, KIC 11026764, KIC 11771760, KIC 2437976, KIC 6370489.

In step (c), we generate half the stellar spectra in the data set with frequency resolution of 4 yrs and the remainder for 3 yrs. When we select samples for training, we linearly interpolate spectra with lower resolution to a higher resolution. This step is crucial, as real data show fluctuations in frequency resolution.

In the final step (d), we avoid class imbalance by drawing all the samples from a uniform prior in parameter space (except inclination angle). The inclination angles are drawn from a uniform distribution in $\sin\iota$. To avoid bias in the parameter prediction, we actively reject correlations among different seismic parameters.   

\begin{deluxetable*}{cccccc}
\tablecaption{Parameter space in different evolutionary stages of giant stars.\label{tab:dataset_table_extended}}
\tablewidth{0pt}
\tablehead{
\colhead{Parameter} & \colhead{Subgiant} & \colhead{Young red giant branch} & \colhead{Old red giant branch} & \colhead{Red clumps} & \colhead{Red giant simulations}\\
}
\startdata
        Range of $\Delta \nu$ &  18-50 $\mu$Hz & 9-18 $\mu$Hz & 6-9 $\mu$Hz & 4.2-12 $\mu$Hz & 1-18.7 $\mu$Hz \\ 
        Range of $\Delta \Pi$ & 60-200s  & 45-150s & 45-110s & 150-500s & 45-500s\\
        Range of q & 0.05-0.5  & 0.05-0.5 & 0.05-0.5 & 0.05-0.75& 0-0.75 \\
        Range of $\epsilon_p$ & 0-1  & 0-1 & 0-1 & 0-1& 0-1. \\
        Range of $\epsilon_g$ & 0-1  & 0-1 & 0-1 & 0-1& 0-1. \\
        Range of $d_{0\ell}$ & 0.005-0.025  & 0.005-0.025 & 0.005-0.025 & 0.005-0.025 & 0.005-0.025 \\
        Range of Core & 0.005-4.0  & 0.005-4.0 & 0.05-3.0 & 0.005-1.0 & 0.005-1.0 \\
        rotation ( in $\mu$Hz) & {} & {} & {} & {} & {} \\
        Range of Envelope & 0.005-0.4  & 0.005-0.4  & 0.005-0.4  & 0.005-0.4 & 0.005-0.4 \\
        rotation ( in $\mu$Hz) & {} & {} & {} & {} & {} \\
        Range of $\iota$ ( in deg) & 0-90  & 0-90 & 0-90 & 0-90 & 0-90\\
        Range of $A_{g}$  & 0.8-1.2 & 0.8-1.2 & 0.8-1.2 & 0.8-1.2 & 0.8-1.2\\
        Range of $B_{g}$ & -2.2 - -1.8 & -2.2 - -1.8 & -2.2 - -1.8 & -2.2 - -1.8 & -2.2 - -1.8\\
        Range of $C_{g}$ & 0-0.5 & 0-0.5 & 0-0.5 & 0-0.5 & 0-0.5 \\
        Range of $A_{\tau}$ & 0.8-1.2 & 0.8-1.2 & 0.8-1.2 & 0.8-1.2 & 0.8-1.2\\
        Range of $B_{\tau}$ & -1.2 - -0.8 & -1.2 - -0.8 & -1.2 - -0.8 & -1.2 - -0.8 & -1.2 - -0.8 \\
        Range of $C_{\tau}$ & 0-0.5 & 0-0.5 & 0-0.5 & 0-0.5 & 0-0.5\\
        Range of $p$ & 1.8-2.2 & 1.8-2.2 & 1.8-2.2 & 1.8-2.2 & 1.8-2.2\\
        Range of $N_{0}$ & 0.1-0.4 & 0.1-0.4 & 0.1-0.4 & 0.1-0.4 & 0.001-40,000\\
        Frequency range used  & 250-1150$\mu$Hz  &  60-262 $\mu$Hz & 22-192 $\mu$Hz & 22-192 $\mu$Hz & 0-250$\mu$Hz \\
        for ML training & {} & {} & {} & {} & {} \\
        Range of Observation & 1065-1460 days & 1065-1460 days & 1065-1460 days & 1065-1460 days & 9-1460 days \\
        time & {} & {} & {} & {} & {} \\
        SNR distribution & 5-70 & 5-70 & 5-70 & 5-70 & 5-150 \\
        \hline
\enddata
\tablecomments{First four column show the parameter space in different evolutionary stages of giant stars.This table shows the range of parameters that were chosen to create different synthetic data sets. This range of parameters is chosen so as to cover the space of published results on \textit{Kepler} Data \cite{Mosser_2015,period_spacing_2,mixed_mode}. Last column shows the parameter space used to produce the red-giant and pure noise simulations. The noise simulations are produced with a 0 signal to noise ratio (SNR).}
\end{deluxetable*}

\subsection{Machine Learning Model}
\label{Machine Learning Model}

Machine learning methods have a common algorithmic approach, namely to train a machine to carry out a task using a training data set. Here, we use a deep neural network, detailed hereafter for the purpose of power spectral analysis.

We want the machine to recognize the non-linear relationship between the normalized spectral data X and the (seismic) parameters Y (in Table \ref{tab:dataset_table}) using deep neural networks $f$, i.e., 
\begin{equation}
    f(\mathrm{X;W}) \approx \mathrm{Y}
    \label{eq_network},
\end{equation}
where rows of X are examples presented to the machine to learn about the parameters, W represents the neural network parameters, and Y is the dependent variable (seismic parameter). Each row of X is a normalized power spectrum. It is obtained by dividing the power spectrum by the maximum power in the used frequency range. X and Y have same number of rows. Rather than predicting real-valued seismic parameters, we first pose a classification problem by categorizing the seismic parameter space using uniformly-spaced bins. This turns each row of Y into a one-hot encoded vector. In one-hot encoding, all the elements of the vector are 0 except the ground truth, which is represented by 1. In this case, the number of columns of Y is the number of bins and a 1 is associated with the ground truth. For example, let a seismic parameter $\theta$ be in range 0 to 50 and the number of bins be 5. Let $i$-th data be generated by a seismic parameter $\theta_i$ of value 37. Then, this target  $\theta_i$ is encoded as [0,0,0,1,0] in the corresponding row of Y. Also, each bin in this method is represented by its mean value (i.e., 5, 15, 25, 35, and 45, respectively, in this example) and a predicted parameter takes only one of these mean values.

For a given normalized spectrum as input, the network outputs a vector of classification scores for the bins. We then apply a so-called softmax function \citep{10.5555/1162264,Goodfellow-et-al-2016}  to convert these scores to probability values and finally apply categorical cross-entropy loss \citep{murphy2013machine, Goodfellow-et-al-2016} on the output probability. We employ the ADAM optimizer \citep{kingma2017adam} for back propagation. Back propagation trains the neural network by adjusting its parameters W such that they minimise the loss function through a feedback loop between the outputs and the inputs. After completing the training, the network learns the approximate marginal probability distribution of the corresponding seismic parameter.
As input, the bin with highest probability ($p_{max}$) is the best-fit seismic parameter whose value is the mean over that bin. We term $p_{max}$ as ``confidence" since it is the confidence score of the predicted seismic parameter. 

The choice of bin size is an important factor in this method. For larger bin sizes, a single representative value fails to capture the larger variety of data and consequently, parameter sensitivity is lost in the prediction. On the other hand, smaller bin sizes indicate superior resolution in parameter space. However, this results in a much larger classification problem that requires enormous amounts of data that well-represent each class. In other words, with limited data, such predictions become unreliable.

\begin{figure*}[!t]
     \centering
     \includegraphics[width=170mm,scale=0.8]{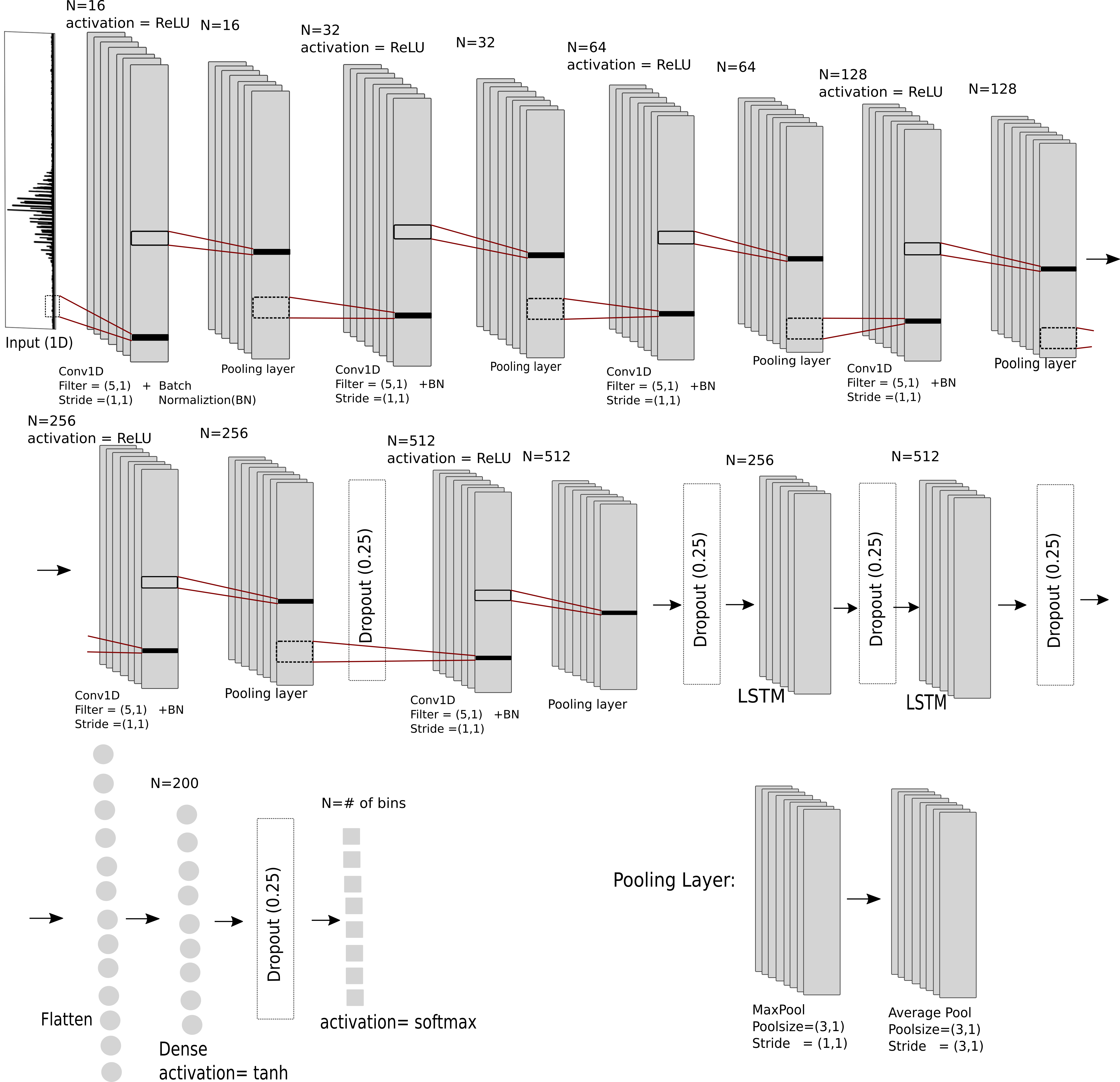}
     \caption{Detailed architecture of neural network used in our machine learning algorithm. The core network consists of 6 CNN layers, 2 LSTM cells and a dense layer. In addition to them, 5 dropout layers with a fraction of 0.25 are used to prevent overfitting. \textit{right-bottom}: This figure describes the components of Pooling layer. This network takes as input the (1D) normalized spectrum and outputs an approximate probability distribution of the subject parameter.}
    \label{fig:ml_network_detail}        
\end{figure*}

Figure \ref{fig:ml_network_detail} shows the detailed architecture of the machine learning network, which is built by these layers. The core network comprises six convolutional layers (conv1D), followed by two long short term memory (LSTM) cells, and one dense layer. This network takes an input normalized power spectrum of length and outputs the probability associated with each bin of seismic parameter. 

\textbf{Computational time}: 
This network has been trained on a single Intel\textsuperscript{\textregistered} Xeon\textsuperscript{\textregistered} Platinum 8280 CPU with 56 cores  using tensorflow, a python based software used for machine learning \citep{tensorflow2015-whitepaper}. The training took $\sim$ 50 node hours for each seismic parameter. For three seismic parameters in four different regimes, we train a separate network with identical architecture. The computational cost for training these 12 networks is 600 node hours (50 node hours each). The trained machine takes $\sim$5 milliseconds of computational time on a single core to predict one parameter on a star. In comparison, estimation of (i) $\Delta \nu$ takes $\mathcal{O}$(min)/star using auto-correlation method (ii)  $\nu_{max}$ takes $\mathcal{O}$(min)/star by fitting a gaussian-envelope (iii) $\Delta \Pi$ takes $\mathcal{O}$(hr)/star using MCMC.

\section{Results on Synthetic data} 
\label{Results_on_Synthetics}

\begin{figure*}
\gridline{\fig{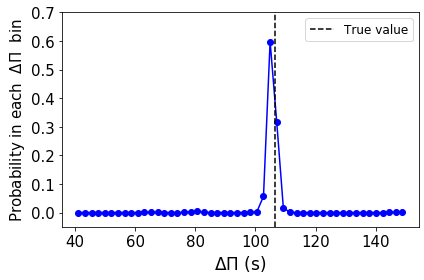}{0.30\textwidth}{(a)}
          \fig{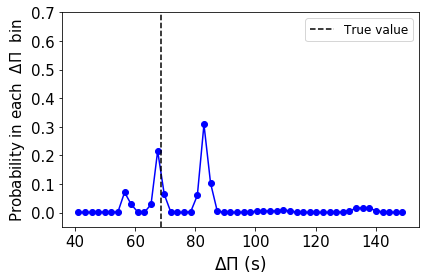}{0.30\textwidth}{(b)}
          \fig{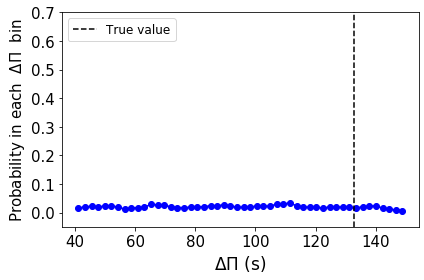}{0.30\textwidth}{(c)}
          }
\caption{Probability distribution of the period-spacing $\Delta \Pi$ for three different synthetic young red giant branch. The true value that generates the spectrum is indicated by the black dashed line. The majority($>$95\%) of the probability distributions are of type (a) and 5\% of the distributions include types (b) and (c).}
\label{fig:prob_dist_diff_distributions}
\end{figure*}

\begin{deluxetable*}{ccccc}
\tablecaption{Choice of $\Delta \nu$, $\nu_{max}$ and $\Delta \Pi$ resolution for different stellar classes.\label{tab:frequency_spacing_resolution}}
\tablewidth{0pt}
\tablehead{
\colhead{Stellar class} & \colhead{$\Delta \nu$ bins ($\mu$Hz)} & \colhead{$\nu_{max}$ bins ($\mu$Hz)} & \colhead{$\Delta \Pi$ bins (s)} \\
}
\startdata
         Subgiants &  [18,18,2],[18.2,18.4], & [229.88,233.89],$\dots$ & [60,62.8],$\dots$,\\
         {} &  $\dots$,[48.8,50] & [952.30,956.31] & [197.2,200]\\
         \hline
         Young red giant branch  &  [9,9.2],[9.2,9.4], & [93.49,95.47],$\dots$, & [45,47.28],$\dots$,\\
         {}  &  $\dots$,[17.8,18] & [251.95,253.93] & [147.72,150]\\
         \hline
         Old red giant branch  &  [6.0,6.1],[6.1,6.2] & [55.18,57.18],$\dots$, & [45,47.16],$\dots$,\\
         {}  &  $\dots$,[8.9,9] & [102.18,104.18] & [107.84,110]\\
         \hline
         Red clump stars & [4.2,4,3],[4.3,4.4], & [34.75,36.73],$\dots$, & [150,157],$\dots$,\\
         {} & $\dots$,[11.9,12] & [147.99,149.97] & [493,500]\\         
         \hline
\enddata

\end{deluxetable*}

For each stellar class given in Table \ref{tab:dataset_table}, we train the machine separately, leaving four different machines for each parameter, where the core networks (CNN-LSTM-dense) are identical but the final layers change in accordance with the parameter and resolution required. Choices for the bins and parameter resolution in each evolutionary regime are given in Table \ref{tab:frequency_spacing_resolution}.

Once trained, the neural network's performance is evaluated on a test data set, which comprises the unseen data. For the preparation of the test data set, we generate 120,000 synthetic stars in various evolutionary stages. The test and training data sets follow the same distribution in parameter space, as described in Table \ref{tab:dataset_table}. We present the results of $\Delta \nu$, $\nu_{max}$, and $\Delta \Pi$ predictions on this data set. 

For every input of a synthetic star's normalized power spectrum, the output of the network is the probability in each bin, from which we construct the approximate probability distribution. These distributions take on varied forms, as demonstrated in Figure \ref{fig:prob_dist_diff_distributions}.

Figure~\ref{fig:prob_dist_diff_distributions} shows the probability distributions of $\Delta\Pi$ in three different stars. For the probability distribution shown in Figure \ref{fig:prob_dist_diff_distributions}(a), the true value which generated the synthetic spectrum matches the neural network prediction with a confidence ($p_{max}$) of 0.6 approximately. More than 95\% of the probability distributions are of the type \ref{fig:prob_dist_diff_distributions}(a) but in the remaining $\sim5\%$ of other cases, we encounter distributions of types \ref{fig:prob_dist_diff_distributions}(b) and \ref{fig:prob_dist_diff_distributions}(c). For the multimodal distribution shown in the Figure \ref{fig:prob_dist_diff_distributions}(b), the true value matches the second peak of the distribution, whereas the distribution in Figure \ref{fig:prob_dist_diff_distributions}(c) is flat. The machine is successful in its first prediction whereas it fails in two other cases. Although the machine is unsuccessful in the second case (Fig. \ref{fig:prob_dist_diff_distributions}(b)), it is possible to test if any of the peaks in the distribution fit the spectrum using forward calculations. Obtaining a rapid estimate of the distribution is valuable since, despite the low confidence results, these could still serve as priors for methods such as Bayesian inference.

\begin{figure*}
\gridline{\fig{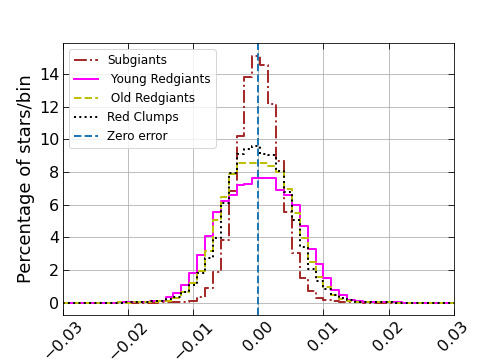}{0.46\textwidth}{(a)}
          \fig{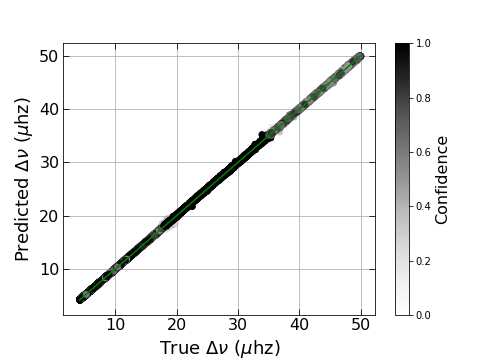}{0.46\textwidth}{(b)}
         }
\caption{Summary of results of $\Delta \nu$ predictions on synthetic data. (a) Relative prediction error in different regimes of synthetic data (legends). The relative error is calculated in reference to the true value that generates the spectrum. The blue-dashed line marks the zero-error. This shows that predictions lie within 1\% of the true values. (b) Predicted values of $\Delta \nu$ at each value of true $\Delta \nu$ for all 120,000 synthetic red giants ($\Delta \nu$:4.2-50$\mu$Hz), across all stellar classes. The color of each point in the plot represents prediction confidence. The green line shows \textit{Predicted} $\Delta \nu$ = \textit{True} $\Delta \nu$. These figures show that $\Delta \nu$ can be inferred within 1.5\% of the ground truth for 99.9\% of stars.}
\label{fig:dnu_results_syn}
\end{figure*}

\begin{figure*}
\gridline{\fig{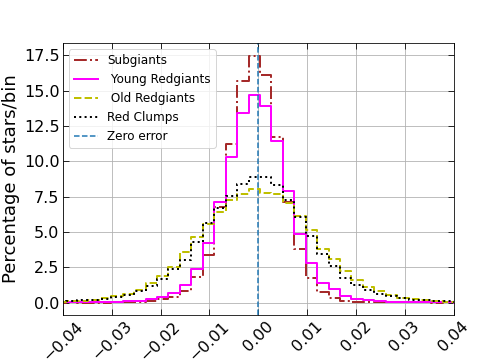}{0.46\textwidth}{(a)}
          \fig{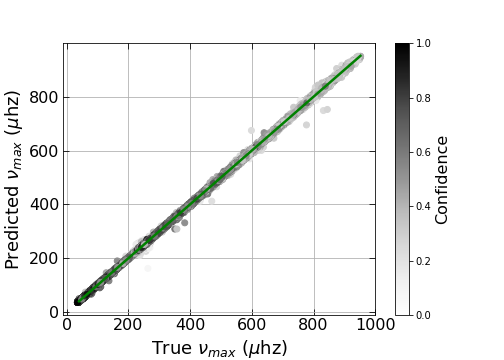}{0.46\textwidth}{(b)}
         }
\caption{Summary of results of $\nu_{max}$ predictions on synthetic data.(a) Relative prediction error in different regimes of synthetic data (legends). Relative error is calculated in reference to the true value that generates the spectrum. The blue-dashed line marks the zero-error. This shows that predictions lie within 2.5\% of true values. (b) Predicted value of $\nu_{max}$ at each value of true $\nu_{max}$ for all 120,000 synthetic red giants, across all stellar classes. The colour of each point in the plot represents prediction confidence. The green line shows \textit{Predicted} $\nu_{max}$ = \textit{True} $\nu_{max}$. These figures show that $\nu_{max}$ can be inferred within 3\% of the ground truth for 99.9\% of stars.}
\label{fig:numax_results_syn}
\end{figure*}

\begin{figure*}
\gridline{\fig{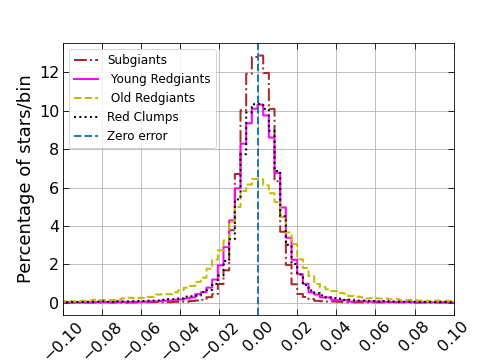}{0.46\textwidth}{(a)}
          \fig{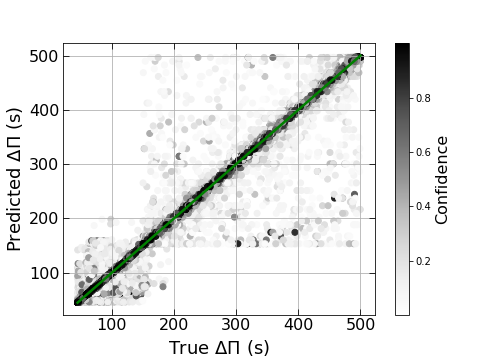}{0.46\textwidth}{(b)}
         }
\caption{Summary of results of $\Delta \Pi$ predictions on synthetic data. (a) Relative prediction error in different regimes of synthetic data (legends). Relative error is calculated in reference to the true value that generates the spectrum. The blue-dashed line marks the zero-error. This shows that predictions lie within 4\% of true values. (b) Predicted value of $\Delta \Pi$ at each value of true $\Delta \Pi$ for all 120,000 synthetic red giants, across all stellar classes. The color of each point in the plot represents prediction confidence. The green line shows \textit{Predicted} $\Delta \Pi$ = \textit{True} $\Delta \Pi$. These figures show that $\Delta \Pi$ can be inferred within 4\% of the ground truth for 99.5\% of stars.}
\label{fig:dp_results_syn}
\end{figure*}

As in Figure~\ref{fig:prob_dist_diff_distributions}, we infer $\Delta \nu$, $\nu_{max}$, and $\Delta \Pi$ from an ensemble of 30,000 synthetic stars in each stellar class of Table \ref{tab:dataset_table}. We present these results in Figures~\ref{fig:dnu_results_syn}, ~\ref{fig:numax_results_syn}, and~\ref{fig:dp_results_syn}. 

Figure~\ref{fig:dnu_results_syn}(a) demonstrates that the machine is able to recover $\Delta \nu$ to within 1\% of the original $\Delta \nu$. Figure \ref{fig:dnu_results_syn}(b) shows the variation of predictions with true values of $\Delta \nu$. It indicates that predictions and true values are highly correlated. Therefore, these figures prove that the machine can identify $p$-mode patterns in synthetic data and predict $\Delta \nu$ accurately. 

Figure~\ref{fig:numax_results_syn}(a) shows that the machine is able to recover $\nu_{max}$ to within 2.5\% of the true $\nu_{max}$. Figure~\ref{fig:numax_results_syn}(b) graphs the variation of predictions with true values of $\nu_{max}$. It demonstrates that predictions and the true values are highly correlated. Therefore, these figures prove that the machine can predict $\nu_{max}$ accurately on synthetic stars. 

Figure~\ref{fig:dp_results_syn}(a) shows that the machine is able to recover $\Delta \Pi$ to within 2.5\%. It also indicates that the machine performs relatively poorly on old red giant branch, where the error is 4\%, compared to the error of $<3\%$ on the other stellar classes. Figure \ref{fig:dp_results_syn}(b) shows the variation of predictions with the true value of $\Delta \Pi$ that generates the power spectrum. It demonstrates that predictions and true values are highly correlated and that the correlation increases with confidence in predictions. Therefore, these results indicate that the machine can identify mixed mode pattern in synthetic data and infer $\Delta \Pi$ accurately.

\begin{figure}[!ht]
     \centering
     \includegraphics[scale=0.5]{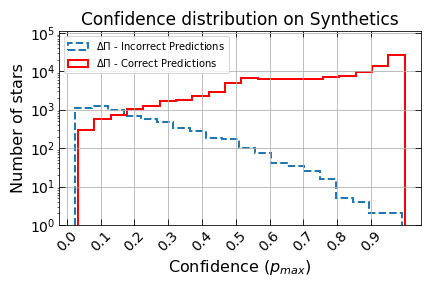}
     \caption{This plot shows the distribution of confidence in $\Delta \Pi$ predictions. It shows the histograms of confidence in \textit{Correct predictions} (Predictions of $\Delta \Pi$ within $<=$5\% of ground truth) and \textit{Incorrect predictions} (Predictions of $\Delta \Pi$  exceeding $>$5\% of ground truth) on an ensemble of 120,000 synthetic stars, across all the stellar classes. The red (blue-dashed) histogram is the confidence distribution in \textit{Correct predictions} (\textit{Incorrect predictions}) predictions. Accuracy is seen to increase with growing confidence.}
     \label{fig:conf_hist_dp}
\end{figure}

Figure~\ref{fig:prob_dist_diff_distributions} shows predictions with three different confidence ($p_{max}$) values. We highlight the importance of confidence in Figure~\ref{fig:conf_hist_dp}, which plots the confidence distributions in correct and incorrect predictions of $\Delta \Pi$ with reference to true values.  If the relative error is less than 5\%, predictions are considered \textit{correct} and otherwise, \textit{incorrect}. Figure \ref{fig:conf_hist_dp} suggests that predictions with higher confidence ($p_{max}$) are more likely to be \textit{correct}. Therefore, confidence acts as an indicator of the accuracy of prediction. In the following subsection, we also show that confidence ($p_{max}$) values produced by the machine are calibrated and represent true likelihood of the prediction.   

\subsection{Calibration test of the Network}
We define two goals for this network: to be accurate and to provide the right estimate of the likelihood. In other words, the confidence estimates provided by the network have to return the correct likelihood/probability. Therefore, the network needs to be calibrated, for which we perform the following test \cite{pmlr-v70-guo17a}.

Consider that the machine returns a value $\hat{Y}$ with confidence $\hat{P}$. If the network is calibrated, it returns the true probability. Therefore,

\begin{equation}
    \mathbf{P}(\hat{Y}=Y|\hat{P}=p) = p.
\end{equation}

Suppose we have $N$ predictions at confidence of $p$: the expected number of accurate predictions is therefore $Np$. The prediction is deemed accurate if the neural network finds the bin corresponding to the ground truth. In this calibration test, we compare the fraction of accurate predictions with confidence values in the range ($p$-$\epsilon$, $p$+$\epsilon$) to confidence $p$ and expect them to be identical.

\begin{figure}[ht!]
     \centering
     \includegraphics[width=85mm]{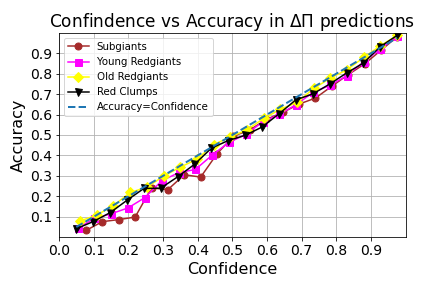}
     \caption{Accuracy as a function of confidence in different regimes of synthetic data (legends). The blue dashes track the \textit{Accuracy = Confidence } line. This shows that accuracy is approximately equal to confidence at each confidence level in different regimes, which  agrees with the calibration test.}
    \label{fig:conf_test_1_dp}      
\end{figure}
The results of the calibration test are presented in Figure \ref{fig:conf_test_1_dp}. The accuracy, shown for the range of confidence values in each stellar class, is approximately equal to confidence at every confidence level and every evolutionary stage. We interpret this result as indicating that the machine has passed the test in all evolutionary stages of giant stars and provides meaningful likelihood values.

\section{}

\begin{figure*}[!ht]
\gridline{\fig{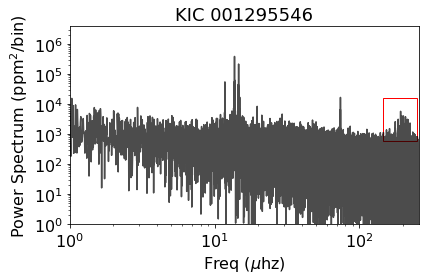}{0.46\textwidth}{(a)}
          \fig{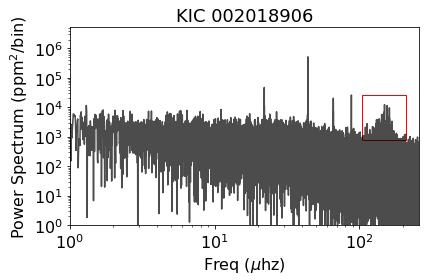}{0.46\textwidth}{(b)}
         }
\caption{(a) Power spectrum of the binary KIC 1295546 where the red-box in the power spectrum highlights the red-giant counterpart. (b) Power spectrum of the binary KIC 2018906 where the red-box in the power spectrum highlights the red-giant counterpart.}
\label{fig:special_systems_b}
\end{figure*}

\startlongtable
\begin{deluxetable*}{cccccccc}
\tablecaption{This table shows the list of new giant stars that are discovered by Machine. It lists the measurements of $\Delta \nu$, $\nu_{max}$ and $\Delta \Pi$. \label{tab:table_redgiants_new}}
\tablehead{
\colhead{} & \colhead{KIC ID} & \colhead{$\Delta \nu$} & \colhead{\it{Error}} & \colhead{$\Delta \Pi$} & \colhead{\it{Error}} & \colhead{$\nu_{max}$} & \colhead{\textit{Error}}\\
\colhead{} & \colhead{} & \colhead{($\mu$Hz)} & \colhead{in $\Delta \nu$ ($\mu$Hz)} & \colhead{(s)} & \colhead{in $\Delta \Pi$} & \colhead{($\mu$Hz)} & \colhead{in $\nu_{max}$ ($\mu$Hz)}\\
}
\startdata
1  &    1569849 &  11.71 &  0.06 &   78.24 &   2.73 &  134.658 &   1.686 \\
2  &    2018906 &  12.10 &  0.04 &   76.33 &   1.86 &  149.159 &   1.015 \\
3  &    2997178 &  15.13 &  0.07 &   85.55 &   0.84 &  191.466 &   1.403 \\
4  &    5396898 &  16.00 &  0.04 &   87.68 &   1.29 &  212.287 &   2.402 \\
5  &    6286155 &  10.89 &  0.04 &   51.05 &   2.62 &  134.898 &   2.226 \\
6  &    6363746 &  12.10 &  0.04 &   41.21 &   2.84 &  156.497 &   1.753 \\
7  &    8144355 &  14.01 &  0.07 &   87.18 &   2.33 &  173.055 &   2.838 \\
8  &    9339382 &  17.09 &  0.08 &   89.61 &   2.75 &  229.104 &  18.333 \\
9  &    9468382 &  12.48 &  0.06 &   81.08 &   1.03 &  148.277 &   1.125 \\
10 &   11081697 &  13.40 &  0.04 &   79.73 &   1.58 &  149.065 &   1.540 \\
11 &    2988153 &   9.18 &  0.06 &   46.92 &  19.23 &  121.573 &   3.628 \\
12 &    4265444 &   9.01 &  0.05 &   82.79 &  13.62 &  100.784 &   1.685 \\
13 &    5557810 &   4.29 &  0.04 &  272.86 &   7.22 &   35.709 &   0.438 \\
14 &    6521537 &   8.20 &  0.04 &   74.26 &  84.40 &   89.784 &   3.405 \\
15 &    9532737 &   6.38 &  0.07 &   92.40 &  10.24 &   60.306 &   3.985 \\
16 &    9594499 &   9.11 &  0.06 &   69.79 &   4.41 &   93.014 &   3.737 \\
17 &    9655198 &   8.08 &  0.07 &  159.56 &  12.55 &   93.164 &   5.264 \\
18 &    9715513 &   9.11 &  0.07 &   75.17 &   8.70 &   96.408 &   5.345 \\
19 &   10399343 &  12.12 &  0.07 &   84.78 &  22.45 &  158.092 &   8.388 \\
20 &   10722800 &  12.71 &  0.07 &   73.14 &  20.48 &  157.912 &  10.766 \\
21 &   12406750 &   7.68 &  0.06 &  289.74 &   3.71 &   51.564 &   0.312 \\
22 &  100000928 &   4.69 &  0.07 &  279.98 &  10.78 &   46.232 &   2.009 \\
23 &    2448669 &   4.30 &  0.04 &  324.85 &  10.48 &   38.383 &   0.436 \\
24 &    3338674 &  10.82 &  0.06 &   83.44 &   1.68 &  122.016 &   1.590 \\
25 &    1295546 &  15.59 &  0.04 &   85.54 &   0.87 &  202.366 &   0.509 \\
\enddata
\end{deluxetable*}

\startlongtable
\begin{deluxetable*}{cccccccc}
\tablecaption{List of 195 redgiant stars detected in \cite{Hon2019}. This table shows the first measurements of $\Delta \nu$ and $\Delta \Pi$ for these stars. \label{tab:table_redgiants_2}}
\tablehead{
\colhead{} & \colhead{KIC ID} & \colhead{$\Delta \nu$} & \colhead{\it{Error}} & \colhead{$\Delta \Pi$} & \colhead{\it{Error}} & \colhead{$\nu_{max}$} & \colhead{\textit{Error}}\\
\colhead{} & \colhead{} & \colhead{($\mu$Hz)} & \colhead{in $\Delta \nu$ ($\mu$Hz)} & \colhead{(s)} & \colhead{in $\Delta \Pi$} & \colhead{($\mu$Hz)} & \colhead{in $\nu_{max}$ ($\mu$Hz)}\\
}
\startdata
1   &   2298039 &   8.80 &  0.04 &   71.81 &    7.71 &   98.807 &   1.846 \\
2   &   2438191 &   8.30 &  0.04 &   71.35 &    5.61 &   92.788 &   1.910 \\
3   &   3113323 &  18.86 &  1.69 &   92.97 &    1.52 &  247.600 &   4.596 \\
4   &   3428926 &   4.61 &  0.10 &  270.35 &    3.92 &   41.357 &   2.159 \\
5   &   3526706 &  15.41 &  0.06 &   92.43 &    9.67 &  200.392 &   5.342 \\
6   &   4039700 &   4.97 &  0.07 &  304.95 &    7.12 &   33.938 &   0.961 \\
7   &   4044655 &  11.39 &  0.04 &  120.51 &   10.27 &  151.091 &   6.673 \\
8   &   4048300 &  14.30 &  0.04 &   81.22 &    0.96 &  185.324 &   1.750 \\
9   &   4156731 &  12.50 &  0.07 &  136.48 &   16.52 &  141.092 &  11.109 \\
10  &   4165911 &   7.20 &  0.04 &   63.27 &    6.72 &   71.242 &   0.833 \\
11  &   4726609 &  12.49 &  0.04 &   81.20 &   15.03 &  151.473 &   3.977 \\
12  &   5007487 &   6.36 &  0.07 &  292.79 &    6.57 &   72.380 &   0.943 \\
13  &   5007766 &   5.70 &  0.04 &  318.31 &    6.19 &   37.632 &   1.110 \\
14  &   5090985 &  18.89 &  0.04 &   92.18 &    1.69 &  256.057 &   1.079 \\
15  &   5111987 &   8.13 &  0.07 &   70.03 &    3.43 &   90.731 &   1.005 \\
16  &   5166009 &   4.91 &  0.06 &  298.41 &   12.73 &   33.485 &   0.607 \\
17  &   5174603 &   5.97 &  0.07 &   64.13 &  206.63 &   58.537 &   0.878 \\
18  &   5391531 &   8.20 &  0.04 &  221.06 &    9.98 &  102.588 &   2.067 \\
19  &   5471035 &  18.31 &  0.08 &   86.30 &    2.71 &  247.283 &   5.123 \\
20  &   5530029 &   9.81 &  0.06 &   71.21 &    4.94 &  110.933 &   0.977 \\
21  &   5535552 &   5.99 &  0.10 &  306.13 &    7.29 &   40.145 &   0.987 \\
22  &   5615905 &   4.83 &  0.07 &  325.13 &    7.82 &   32.756 &   0.917 \\
23  &   5737554 &   7.81 &  0.06 &  250.42 &   12.52 &   95.548 &   2.317 \\
24  &   5780414 &  15.00 &  0.04 &   85.40 &    0.99 &  195.646 &   2.228 \\
25  &   5787662 &   7.20 &  0.04 &  251.55 &    6.10 &   89.895 &   2.285 \\
26  &   5869582 &  14.81 &  0.06 &   84.94 &    1.52 &  191.391 &   1.127 \\
27  &   5943771 &  13.27 &  0.07 &   90.31 &   11.43 &  205.890 &   7.449 \\
28  &   5975275 &   9.43 &  0.07 &   60.37 &   24.98 &  116.457 &   1.196 \\
29  &   6032981 &   5.18 &  0.06 &  312.75 &    6.69 &   35.489 &   0.935 \\
30  &   6038665 &   9.69 &  0.04 &   70.01 &    3.46 &  106.237 &   0.798 \\
31  &   6289516 &  15.13 &  0.07 &   85.60 &    0.91 &  194.802 &   1.287 \\
32  &   6367082 &  10.53 &  0.07 &   77.13 &    1.65 &  119.330 &   0.762 \\
33  &   6367296 &  17.03 &  0.07 &   90.90 &    1.55 &  221.108 &   1.247 \\
34  &   6368892 &  16.31 &  0.08 &   92.49 &    1.12 &  249.332 &   5.808 \\
35  &   6526965 &   9.76 &  0.07 &   80.47 &    9.75 &  113.904 &   2.030 \\
36  &   6763283 &   4.50 &  0.04 &  313.08 &    6.85 &   30.667 &   0.944 \\
37  &   6861592 &   4.71 &  0.07 &  311.57 &    5.84 &   31.719 &   0.354 \\
38  &   7035674 &  18.39 &  0.08 &   81.71 &    1.60 &  251.551 &   4.239 \\
39  &   7272363 &  18.87 &  0.08 &   93.37 &    2.62 &  255.032 &   2.130 \\
40  &   7347185 &  13.08 &  0.07 &   79.67 &    2.31 &  140.094 &   4.528 \\
41  &   7417006 &  16.23 &  0.07 &   88.16 &    1.21 &  204.289 &   2.855 \\
42  &   7429268 &  12.12 &  0.07 &   75.83 &    6.79 &  141.845 &   1.483 \\
43  &   7433931 &   4.39 &  0.04 &  289.09 &    5.88 &   30.374 &   1.225 \\
44  &   7448275 &  16.74 &  0.07 &   88.13 &    1.31 &  226.119 &  15.606 \\
45  &   7581425 &  16.12 &  0.07 &   83.73 &    1.39 &  220.506 &   1.733 \\
46  &   7593204 &  18.43 &  0.08 &   90.88 &    2.13 &  254.106 &   2.997 \\
47  &   7596238 &  17.61 &  0.07 &   88.07 &    0.97 &  235.260 &   3.858 \\
48  &   7679919 &  15.59 &  0.07 &   87.02 &    1.74 &  203.868 &   2.474 \\
49  &   7801777 &  18.89 &  0.68 &   98.43 &    2.17 &  254.944 &   1.820 \\
50  &   7810482 &  18.43 &  0.07 &   89.00 &    1.60 &  251.966 &   3.573 \\
51  &   7877928 &  14.70 &  0.04 &   81.58 &    1.56 &  179.271 &   1.391 \\
52  &   7878393 &  10.27 &  0.07 &   78.12 &    1.52 &  116.168 &   0.878 \\
53  &   8075941 &  10.62 &  0.07 &   70.48 &   14.53 &  135.067 &   1.754 \\
54  &   8095479 &  17.42 &  0.07 &   86.05 &    9.68 &  248.385 &   5.940 \\
55  &   8159089 &  17.02 &  0.07 &   87.89 &    0.81 &  218.386 &   1.980 \\
56  &   8176747 &  18.89 &  0.04 &   92.34 &    1.23 &  255.468 &   1.525 \\
57  &   8212119 &   9.01 &  0.06 &   69.47 &    7.33 &   98.241 &   1.988 \\
58  &   8219710 &  17.22 &  0.06 &   90.00 &    1.39 &  216.067 &   4.401 \\
59  &   8222873 &   4.38 &  0.06 &  318.73 &    7.90 &   32.930 &   0.213 \\
60  &   8329820 &  14.13 &  0.07 &   82.40 &    1.73 &  186.196 &   2.920 \\
61  &   8346067 &  14.44 &  0.07 &   82.95 &    1.27 &  187.541 &   2.095 \\
62  &   8350593 &   4.69 &  0.07 &  293.37 &    9.07 &   31.979 &   0.898 \\
63  &   8416927 &  14.64 &  0.07 &   83.21 &    0.90 &  175.542 &   1.333 \\
64  &   8417929 &   9.50 &  0.04 &   70.01 &    3.41 &  103.704 &   0.838 \\
65  &   8480342 &  17.59 &  0.06 &   88.49 &    1.56 &  233.352 &   7.198 \\
66  &   8494649 &  16.69 &  0.06 &   85.76 &    0.95 &  221.276 &   3.357 \\
67  &   8509950 &  18.39 &  0.08 &   86.43 &    1.71 &  251.933 &   3.495 \\
68  &   8609645 &  17.12 &  0.07 &   86.96 &    1.56 &  231.742 &   1.859 \\
69  &   8611967 &  13.09 &  0.06 &   84.80 &    2.77 &  148.676 &   2.861 \\
70  &   8675208 &   4.30 &  0.08 &  331.88 &    7.28 &   30.430 &   0.904 \\
71  &   8802225 &   4.49 &  0.05 &  324.20 &    8.50 &   31.272 &   0.651 \\
72  &   8804749 &  15.96 &  0.07 &   83.37 &    0.83 &  219.808 &   1.893 \\
73  &   8874261 &  14.70 &  0.04 &   90.32 &    0.89 &  204.030 &   3.255 \\
74  &   9020498 &  18.90 &  0.04 &   95.06 &    1.47 &  254.856 &   2.036 \\
75  &   9080204 &  13.40 &  0.04 &   79.18 &    1.44 &  167.390 &   1.290 \\
76  &   9145612 &  18.89 &  0.04 &   92.38 &    2.40 &  255.194 &   1.755 \\
77  &   9209074 &  14.20 &  0.06 &   83.27 &    0.88 &  173.433 &   1.027 \\
78  &   9216911 &  16.09 &  0.05 &   88.31 &    1.36 &  220.057 &   3.090 \\
79  &   9225884 &  16.81 &  0.06 &   86.47 &    1.81 &  250.505 &   4.578 \\
80  &   9267669 &   8.71 &  0.06 &   70.00 &    3.41 &   88.279 &   0.765 \\
81  &   9272024 &   4.44 &  0.08 &  241.76 &    7.19 &   30.816 &   0.612 \\
82  &   9391471 &   4.49 &  0.07 &  321.69 &    7.40 &   29.893 &   0.823 \\
83  &   9480210 &   4.56 &  0.09 &  276.65 &   13.30 &   31.396 &   0.938 \\
84  &   9518306 &  18.89 &  0.04 &   90.24 &    1.62 &  251.648 &   3.717 \\
85  &   9579611 &   5.72 &  0.07 &   54.60 &   15.07 &   56.080 &   0.623 \\
86  &   9640352 &  17.10 &  0.05 &   87.07 &    1.56 &  230.992 &   1.617 \\
87  &   9691704 &   4.77 &  0.08 &  306.79 &   84.17 &   33.050 &   0.687 \\
88  &   9697265 &  13.88 &  0.07 &   83.45 &    0.89 &  172.115 &   0.826 \\
89  &   9762744 &   4.57 &  0.07 &  332.08 &    6.62 &   31.564 &   0.986 \\
90  &   9815168 &  11.80 &  0.04 &   80.80 &    0.94 &  132.656 &   1.069 \\
91  &   9936758 &  16.90 &  0.08 &  100.23 &    2.12 &  244.552 &  11.197 \\
92  &  10004898 &  17.61 &  0.05 &   89.59 &    1.50 &  226.014 &   2.146 \\
93  &  10091105 &  15.34 &  0.08 &   74.81 &   30.56 &  205.892 &   7.217 \\
94  &  10120966 &  13.20 &  0.04 &   42.12 &    3.17 &  165.404 &   4.233 \\
95  &  10159347 &  12.31 &  0.07 &   75.71 &    8.41 &  160.241 &   2.125 \\
96  &  10264259 &  15.51 &  0.06 &   83.90 &    1.79 &  199.076 &   5.008 \\
97  &  10331512 &  11.28 &  0.06 &   74.48 &    1.33 &  119.343 &   1.132 \\
98  &  10340388 &   9.98 &  0.07 &   75.34 &    1.57 &  103.609 &   0.939 \\
99  &  10449265 &  15.40 &  0.05 &   86.24 &    1.51 &  187.311 &   1.696 \\
100 &  10477733 &  14.91 &  0.04 &   85.71 &    0.88 &  184.466 &   1.185 \\
101 &  10549925 &  12.77 &  0.07 &   70.25 &    4.88 &  155.268 &   3.760 \\
102 &  10793654 &  10.31 &  0.05 &   78.68 &    1.02 &  113.355 &   1.146 \\
103 &  10866844 &   7.19 &  0.06 &  239.34 &    3.88 &   88.986 &   1.208 \\
104 &  10877341 &   9.63 &  0.07 &   83.14 &   15.13 &  110.030 &   1.998 \\
105 &  11018628 &   8.59 &  0.04 &   68.11 &    6.14 &  110.657 &   1.907 \\
106 &  11027938 &  10.21 &  0.04 &   75.99 &    2.08 &  114.429 &   1.181 \\
107 &  11028153 &  16.81 &  0.04 &   89.59 &    1.49 &  212.987 &   0.793 \\
108 &  11076347 &   5.88 &  0.06 &  313.70 &    7.27 &   39.622 &   0.764 \\
109 &  11141326 &  18.40 &  0.08 &   92.26 &    1.06 &  251.970 &   3.913 \\
110 &  11189395 &   4.27 &  0.07 &  273.18 &    8.44 &   28.682 &   1.064 \\
111 &  11241086 &   9.17 &  0.07 &   74.41 &    2.67 &   94.740 &   0.930 \\
112 &  11242390 &   4.85 &  0.07 &  287.24 &    6.95 &   31.959 &   0.959 \\
113 &  11296612 &  15.33 &  0.07 &   84.03 &    1.54 &  197.141 &   1.185 \\
114 &  11305436 &  14.11 &  0.06 &   90.02 &    1.19 &  166.792 &   4.612 \\
115 &  11305445 &  18.90 &  0.04 &   92.85 &    1.79 &  249.401 &   4.494 \\
116 &  11397467 &  18.39 &  0.08 &   90.52 &    1.29 &  241.183 &   3.754 \\
117 &  11408719 &  15.31 &  0.04 &   83.68 &    1.31 &  197.865 &   1.727 \\
118 &  11409098 &  16.83 &  0.07 &   87.92 &    0.82 &  212.923 &   2.059 \\
119 &  11413789 &  14.63 &  0.07 &   80.96 &    0.85 &  196.247 &   2.751 \\
120 &  11450209 &  12.37 &  0.07 &   81.45 &    8.04 &  145.440 &   1.727 \\
121 &  11457312 &  16.13 &  0.07 &   88.19 &    1.25 &  205.849 &   1.553 \\
122 &  11466152 &  14.08 &  0.07 &   80.91 &    0.87 &  180.450 &   1.971 \\
123 &  11551196 &   5.61 &  0.04 &  302.18 &    6.52 &   37.391 &   1.132 \\
124 &  11654022 &   5.00 &  0.04 &  319.94 &    4.23 &   35.854 &   0.731 \\
125 &  11752484 &   8.68 &  0.06 &   70.00 &    3.41 &   87.196 &   0.806 \\
126 &  11804004 &  15.07 &  0.05 &   79.75 &    3.90 &  194.893 &   5.780 \\
127 &  11805449 &  11.51 &  0.06 &   79.59 &    3.68 &  129.567 &   0.936 \\
128 &  11968543 &  10.89 &  0.04 &   78.31 &    1.34 &  115.417 &   0.688 \\
129 &  12007329 &   5.35 &  0.07 &  320.49 &    4.01 &   35.934 &   1.171 \\
130 &  12011307 &  15.58 &  0.07 &   87.57 &    1.29 &  199.101 &   1.890 \\
131 &  12058612 &  15.35 &  0.07 &   85.65 &    0.82 &  194.091 &   1.511 \\
132 &  12069521 &  12.03 &  0.07 &   83.06 &    1.96 &  143.560 &   1.433 \\
133 &  12110876 &  12.05 &  0.07 &   80.88 &    2.75 &  144.192 &   1.045 \\
134 &  12120211 &  11.48 &  0.06 &   76.90 &    5.47 &  129.986 &   1.456 \\
135 &  12120246 &  13.30 &  0.04 &   82.82 &    1.47 &  168.124 &   1.174 \\
136 &  12120409 &  11.40 &  0.04 &   80.76 &    2.68 &  138.761 &   1.289 \\
137 &  12164458 &  17.11 &  0.07 &   90.20 &    0.94 &  227.113 &   3.212 \\
138 &  12164811 &  13.19 &  0.03 &   79.83 &    3.59 &  157.495 &   1.488 \\
139 &  12167756 &  18.85 &  0.09 &   92.85 &    1.62 &  255.142 &   1.548 \\
140 &  12169042 &  16.13 &  0.08 &   87.89 &    0.88 &  205.099 &   1.604 \\
141 &  12302267 &   5.39 &  0.04 &  329.06 &    5.23 &   36.195 &   0.828 \\
142 &  12315908 &  15.73 &  0.07 &   85.57 &    0.98 &  203.615 &   3.470 \\
143 &  12405226 &  16.31 &  0.06 &   86.80 &    1.61 &  214.643 &   3.127 \\
144 &  12418457 &   8.59 &  0.04 &  158.91 &   13.55 &  107.434 &   3.565 \\
145 &   1164571 &  16.18 &  0.07 &   87.79 &    1.07 &  214.181 &   3.793 \\
146 &   2299465 &   8.80 &  0.04 &   70.11 &    3.48 &   98.269 &   1.122 \\
147 &   2424949 &   5.18 &  0.09 &  259.32 &    5.64 &   36.322 &   1.106 \\
148 &   4269116 &   4.95 &  0.08 &  292.64 &    6.88 &   31.882 &   0.868 \\
149 &   5024582 &   4.78 &  0.06 &  327.08 &    7.88 &   45.946 &   0.829 \\
150 &   5528710 &   7.32 &  0.07 &   67.46 &    6.46 &   71.415 &   0.690 \\
151 &   5723895 &   4.38 &  0.07 &  103.26 &    2.54 &   39.777 &   0.903 \\
152 &   6116549 &   4.41 &  0.04 &   58.67 &   11.43 &   36.335 &   0.471 \\
153 &   6124426 &  15.99 &  0.06 &   85.75 &    0.91 &  200.696 &   3.491 \\
154 &   7267121 &   8.19 &  0.04 &   70.36 &    6.87 &   84.752 &   1.888 \\
155 &   7288263 &   4.31 &  0.04 &  258.82 &    8.39 &   35.979 &   1.782 \\
156 &   7840541 &   6.78 &  0.06 &   68.98 &    4.38 &   70.275 &   1.649 \\
157 &   7849945 &   4.79 &  0.09 &  289.39 &   10.78 &   43.632 &   2.387 \\
158 &   7935781 &  17.31 &  0.08 &   89.93 &    1.79 &  236.028 &  12.473 \\
159 &   7954696 &   5.91 &  0.07 &  322.51 &    7.34 &   37.539 &   0.575 \\
160 &   8043592 &  10.23 &  0.08 &   79.19 &    4.93 &  111.361 &   3.570 \\
161 &   8197192 &   4.58 &  0.07 &  314.07 &    7.33 &   43.976 &   0.796 \\
162 &   8650209 &   4.53 &  0.07 &  269.68 &    7.37 &   41.757 &   1.956 \\
163 &   9640172 &   5.26 &  0.07 &  319.80 &    3.95 &   36.633 &   0.531 \\
164 &   9959245 &   8.18 &  0.07 &   72.66 &    6.88 &   80.578 &   1.619 \\
165 &  11241343 &   6.43 &  0.08 &  281.26 &    5.33 &   46.544 &   0.832 \\
166 &  11450313 &   4.80 &  0.05 &  278.49 &    5.97 &   32.213 &   1.223 \\
167 &  12256511 &   6.89 &  0.06 &   60.10 &    3.79 &   71.954 &   2.372 \\
168 &   5705767 &  15.42 &  0.07 &   86.42 &    1.58 &  191.241 &   0.845 \\
169 &   6526898 &  14.22 &  0.06 &   80.77 &    1.51 &  186.010 &   0.728 \\
170 &   6619621 &   4.90 &  0.05 &  313.48 &   11.39 &   47.655 &   0.751 \\
171 &   6952065 &  13.90 &  0.05 &   80.64 &    2.00 &  174.901 &   0.609 \\
172 &   7102376 &   5.03 &  0.07 &  319.50 &    4.40 &   34.016 &   0.374 \\
173 &   7190608 &  15.91 &  0.07 &   88.19 &    1.17 &  210.497 &   0.653 \\
174 &   7428429 &  18.90 &  0.04 &   91.81 &    2.24 &  256.450 &   0.530 \\
175 &   7659432 &   5.20 &  0.04 &  167.69 &    8.02 &   34.296 &   0.303 \\
176 &   7802947 &  18.52 &  0.08 &   89.88 &    1.69 &  242.786 &   1.045 \\
177 &   7887065 &  12.10 &  0.04 &   82.70 &    2.48 &  143.452 &   0.737 \\
178 &   7944463 &   9.60 &  0.04 &   67.25 &    0.86 &   98.153 &   0.449 \\
179 &   8099642 &   9.71 &  0.06 &   70.94 &    4.10 &  110.229 &   0.313 \\
180 &   8284425 &   4.65 &  0.07 &  318.74 &    7.44 &   32.186 &   0.493 \\
181 &   8482229 &  15.71 &  0.06 &   85.91 &    1.21 &  201.803 &   0.618 \\
182 &   8560128 &  14.14 &  0.08 &   71.79 &    0.81 &  195.048 &   0.826 \\
183 &   8605198 &   9.64 &  0.07 &   72.81 &    6.83 &  110.241 &   0.556 \\
184 &   9012209 &   7.62 &  0.07 &   67.47 &    7.80 &   79.327 &   0.276 \\
185 &   9087806 &   7.70 &  0.04 &  182.78 &    6.65 &   95.768 &   0.588 \\
186 &   9214725 &  17.48 &  0.07 &   88.14 &    1.15 &  228.066 &   1.125 \\
187 &   9489955 &   7.08 &  0.07 &  289.89 &    3.84 &   51.508 &   0.194 \\
188 &  10280410 &  16.30 &  0.04 &   87.80 &    0.92 &  213.003 &   0.736 \\
189 &  11818536 &  14.01 &  0.06 &   99.07 &   17.45 &  170.889 &   0.968 \\
190 &  11858309 &   5.09 &  0.05 &  270.88 &   11.43 &   37.593 &   0.044 \\
191 &  12115826 &  15.31 &  0.07 &   83.40 &    0.90 &  201.553 &   0.922 \\
192 &  12504765 &   4.78 &  0.07 &  331.19 &    5.33 &   35.318 &   0.026 \\
193 &   5385245 &   9.08 &  0.06 &  101.58 &    2.83 &  120.754 &   0.820 \\
194 &   6753216 &   4.63 &  0.07 &  304.02 &   10.50 &   42.962 &   0.122 \\
195 &   1161491 &   4.33 &  0.07 &  297.73 &    6.49 &   37.555 &   0.914 \\
\enddata
\end{deluxetable*}

\end{document}